\documentclass[a4paper,11pt]{article}

\usepackage{setup}
\usepackage{jheppub}
\usepackage{todonotes}
\usepackage{placeins}
\usepackage{longtable}

\usepackage{cite}
\usepackage{ifthen}
\usepackage{subcaption}
\usepackage{siunitx}
\usepackage{enumitem}
\usepackage{booktabs}
\usepackage{bookmark}

\usepackage{cleveref}
\DeclareSIUnit\fb{\femto\barn}

\def\lapprox{\lower .7ex\hbox{$\;\stackrel{\textstyle <}{\sim}\;$}}
\def\gapprox{\lower .7ex\hbox{$\;\stackrel{\textstyle >}{\sim}\;$}}
\definecolor{lightgray}{HTML}{A6A39A}
\definecolor{darkgray}{HTML}{504E48}
\definecolor{silver}{HTML}{E0DFDE}
\definecolor{brown}{HTML}{5F4541}
\definecolor{beige}{HTML}{DCCCAC}
\definecolor{green}{HTML}{345F53}
\definecolor{yellow}{HTML}{F6B65A}
\definecolor{blue}{HTML}{568BCF}
\definecolor{red}{HTML}{AE1932}
\definecolor{orange}{HTML}{D16F15}

\makeatletter
\newcommand{\myitem}[1]{%
	\item[#1]\protected@edef\@currentlabel{#1}%
}
\makeatother

\preprint{{\raggedleft%
IPPP/25/53, ZU-TH 52/25, MCNET-25-17 \\
}}

\title{Precise Predictions for Event Shapes in Hadronic Higgs Decays}
\author[a]{Elliot~Fox,}
\author[b,c]{Aude~Gehrmann--De Ridder,}
\author[c]{Thomas~Gehrmann,}
\author[a]{Nigel~Glover,}
\author[a]{Matteo~Marcoli,}
\author[d]{Christian~T.~Preuss}

\affiliation[a]{Institute for Particle Physics Phenomenology, Department of Physics, University of Durham, Durham, DH1 3LE, UK}
\affiliation[b]{Institute for Theoretical Physics, ETH, 8093 Z{\"u}rich, Switzerland}
\affiliation[c]{Physik-Institut, Universit{\"a}t Z{\"u}rich, 8057 Z{\"u}rich, Switzerland}
\affiliation[d]{Institut f{\"u}r Theoretische Physik, Georg-August-Universit{\"a}t G{\"o}ttingen, 37077 G{\"o}ttingen, Germany}

\emailAdd{elliot.fox@durham.ac.uk}
\emailAdd{thomas.gehrmann@uzh.ch}
\emailAdd{gehra@phys.ethz.ch}
\emailAdd{e.w.n.glover@durham.ac.uk}
\emailAdd{matteo.marcoli@durham.ac.uk}
\emailAdd{christian.preuss@uni-goettingen.de}

\abstract{
  We present NNLO QCD predictions for a wide range of event-shape observables in hadronic Higgs decays, taking into account the two dominant decay modes $H\to gg$ and $H\to b\bar{b}$.
  Specifically, we consider the six classical event shapes thrust, heavy jet mass, $C$-parameter, total and wide jet broadening, and the three-jet resolution $y_{23}$ in the Durham algorithm.
  We also present results for the soft-drop variant of thrust.
  Decays of the Higgs boson to two gluons are treated in the heavy-top limit, whereas decays to a bottom-quark pair are mediated by a non-vanishing Yukawa coupling, despite considering kinematically massless quarks.
  Our results highlight the importance of NNLO QCD corrections in the calculation of event-shape observables and provide means to quantify the intrinsic difference between the two Higgs decay modes.
}

\begin{document}
\maketitle
\flushbottom

\section{Introduction}
\label{sec:intro}

Event-shape variables are computed from the 
momenta of all hadrons in the  final state of a high-energy particle collision. 
The resulting differential event-shape distributions 
 constitute a class of observables with particular sensitivity to the dynamics of Quantum Chromodynamics (QCD) 
 at high energies. They are complementary to jet production cross sections and have played
 a crucial role in establishing and validating the structure of QCD in 
hadronic final states through electron-positron annihilation at LEP \cite{ALEPH:1990iba,OPAL:1990xiz,L3:1992btq,DELPHI:1999vbd,DELPHI:2003yqh}.
In many cases, analyses in hadronic $Z$-boson and off-shell photon decays have been carried out for three-jet event shapes, which are non-vanishing only in configurations which contain at least three particles.
For these observables, experimental efforts have been, and still are, complemented by a range of theoretical precision calculations, spanning fixed-order predictions at next-to-next-to-leading order in the strong coupling \cite{Gehrmann-DeRidder:2007vsv,Gehrmann-DeRidder:2009fgd,Gehrmann-DeRidder:2014hxk,Weinzierl:2009nz,Weinzierl:2009ms,Weinzierl:2009yz,DelDuca:2016ily,Kardos:2018kth}, the resummation of logarithmically enhanced terms \cite{Becher:2011pf,Becher:2012qc,Balsiger:2019tne,Hoang:2014wka,Banfi:2014sua,Banfi:2018mcq,Bhattacharya:2022dtm,Bhattacharya:2023qet}, and, more recently, the determination of power corrections to describe non-perturbative contributions \cite{Gehrmann:2010uax,Luisoni:2020efy,Caola:2021kzt,Caola:2022vea,Agarwal:2023fdk}.

Following the discovery of the Higgs boson by the ATLAS \cite{ATLAS:2012yve} and CMS \cite{CMS:2012qbp} experiments at the Large Hadron Collider (LHC), a new electron-positron collider, as for instance the CERN FCC-ee~\cite{Abada:2019zxq}, the CEPC~\cite{CEPCStudyGroup:2018ghi}, or the ILC~\cite{ILC:2013jhg}, presents one of the most promising avenues for precision Higgs measurements. 
These colliders would operate in a ``Higgs-factory'' mode, with a centre-of-mass energy sufficient to copiously produce Higgs bosons in a relatively clean experimental environment without contamination from initial-state QCD radiation and multi-parton interactions.
This will allow in particular for precision studies of the hadronic Higgs boson decay modes, which will rely on a close 
interplay between experimental measurements and theory predictions.  
Our work aims to improve the quantitative precision of these predictions by computing 
the NNLO QCD corrections to event-shape distributions in hadronic Higgs boson decays.
Besides providing theoretical predictions to confront future experimental data, our work also connects to the topic of quark-gluon discrimination, where hadronic Higgs decays are often exploited as proxies for jet-flavour studies~\cite{Banfi:2006hf,Komiske:2018vkc,Caletti:2021oor,Caletti:2022glq,Caletti:2022hnc,Czakon:2022wam,Gauld:2022lem,Caola:2023wpj,Andersen:2024czj,Behring:2025ilo}.

\begin{figure}[t]
  \centering
  \includegraphics[width=0.9\textwidth]{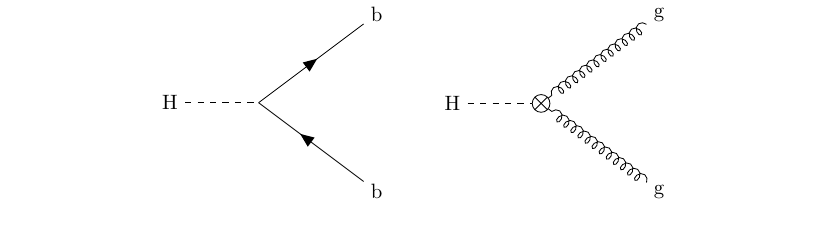}
  \caption{Hadronic Higgs decay categories: $H\to b\bar{b}$ with a Yukawa coupling  (left) and $H\to gg$ via an effective coupling (right).}
  \label{fig:diagH2jLO}
\end{figure}

The hadronic decays of Higgs bosons primarily proceed via the Yukawa-induced decay to a bottom-quark pair
and the loop-induced decay to two gluons, see Figure~\ref{fig:diagH2jLO}.
The former, $H\to b\bar{b}$ decay, has been observed by both ATLAS~\cite{Aaboud:2018zhk} and CMS~\cite{Sirunyan:2018kst} at the LHC, enabled by a measurement in the relatively clean $VH$ production channel.
A similar measurement of the $H\to gg$ decay at the LHC remains elusive, due the large irreducible QCD background in hadron-collider environments.
The inclusive widths of the two hadronic decay channels are known up to fourth order in the strong coupling for $H \to b\bar{b}$ decays (with massless quarks) \cite{Gorishnii:1990zu,Gorishnii:1991zr,Kataev:1993be,Surguladze:1994gc,Larin:1995sq,Chetyrkin:1995pd,Chetyrkin:1996sr,Baikov:2005rw} and for the $H\to gg$ channel \cite{Spira:1995rr,Chetyrkin:1997iv,Baikov:2006ch,Davies:2017xsp,Herzog:2017dtz}
in the limit of infinitely heavy top quark mass~\cite{Wilczek:1977zn,Shifman:1978zn,Inami:1982xt}. 
In both cases, first-order electroweak corrections are available \cite{Fleischer:1980ub,Bardin:1990zj,Dabelstein:1991ky,Kniehl:1991ze,Aglietti:2004nj,Degrassi:2004mx,Actis:2008ug,Aglietti:2004ki}.
A fully-differential calculation of the $H\to b\bar{b}$ decay rate at third order in the strong coupling has been performed in \cite{Mondini:2019gid} and the two-jet rate in the dominant hadronic Higgs decay channels has recently been calculated to third order in QCD in \cite{Fox:2025cuz}.

In the present work, we focus on the
 calculation of  distributions in the  six classical event-shape observables~\cite{Jones:2003yv}.
In the context of hadronic Higgs decays, predictions for these \cite{Coloretti:2022jcl} and related three-jet event shapes \cite{Gao:2020vyx}, four-jet event shapes \cite{Gehrmann-DeRidder:2023uld}, and flavoured observables \cite{CampilloAveleira:2024fll} have previously been calculated at NLO QCD. Employing the designer antenna-subtraction framework~\cite{Braun-White:2023sgd,Braun-White:2023zwd}, and in particular generalised antenna functions~\cite{Fox:2024bfp}, in the \nnlojet \cite{NNLOJET:2025rno} parton-level Monte Carlo event generator, we compute numerical predictions for the 
distributions in thrust~\cite{Brandt:1964sa,Farhi:1977sg}, heavy jet mass~\cite{Clavelli:1981yh}, $C$-parameter~\cite{Parisi:1978eg,Donoghue:1979vi}, wide and total jet broadening~\cite{Rakow:1981qn}, and 
Durham three-jet resolution~\cite{Catani:1991hj} at NNLO QCD accuracy in hadronic Higgs decays.
In case of the thrust observable, we consider both its standard form as well as its definition in conjunction with the soft-drop algorithm \cite{Marzani:2019evv} for three values of the soft-drop parameter $\beta = 0,1,2$.
We perform the calculation in an effective field theory in the heavy-top limit with kinematically massless quarks and a non-vanishing Yukawa coupling for the $b$-quark.
Under these assumptions, all interferences between the $H\to b\bar{b}$ and $H\to gg$ decay channels vanish, allowing us to treat the two decays as separate processes with separate higher-order corrections.
In this way, we will be able to determine qualitative differences between quark-initiated and gluon-initiated QCD radiation.

This paper is structured as follows.
In Section~\ref{sec:framework}, we detail the theoretical framework for our calculation 
and define all observables relevant in this work.
Numerical results are presented in Section~\ref{sec:results}, which also contains an outline of our numerical setup.
We summarize our findings in Section~\ref{sec:conclusion}.

\section{Theoretical Framework}
\label{sec:framework}
Our calculation is based on an effective field theory in which the Higgs boson couples to gluons via an effective $Hgg$ coupling~\cite{Wilczek:1977zn,Shifman:1978zn,Inami:1982xt} and to kinematically massless $b$-quarks via a non-vanishing Yukawa coupling.
In this setup, the \LO inclusive hadronic decay widths are given as
\begin{equation}
  \GammaHbb^{(0)} = \frac{y_b^2(\muR)m_\PH\NC}{8\uppi} \,, \quad \GammaHgg^{(0)} = \frac{\lambda_0^2(\muR)m_H^3(\NC^2-1)}{64\uppi} \, ,
  \label{eq:ratesLO}
\end{equation}
where $m_H$ is the Higgs mass.
The couplings to the Higgs boson are determined by the Yukawa coupling of the $b$-quark, $y_b$, and the \LO effective $Hgg$ coupling in the heavy-top limit, $\lambda_0$.
In terms of the Fermi constant $\GF$, these are given by
\begin{equation}\label{eq:couplings}
  y_b^2(\muR) = m_b^2(\muR)\sqrt{2}\GF\,, \quad \lambda_0^2(\muR) = \frac{\alphas^2(\muR)\sqrt{2}\GF}{9\uppi^2}\,.
\end{equation}
We renormalise both the Yukawa coupling and the effective Higgs-gluon coupling in the $\overline{\text{MS}}$ scheme and evaluate these quantities at $\mu_R$ with $N_F=5$.
The running of the $b$-quark and $t$-quark mass $m_b$ and $m_t$, entering the Yukawa coupling and higher-order corrections of the top-quark Wilson coefficient, respectively, is taken into account using the results of \cite{Vermaseren:1997fq}.

The partonic channels that contribute to our calculation up to NNLO are summarised in Table~\ref{tab:channels}.
We implement all matrix elements in analytical form.
Tree-level matrix elements with up to five partons have been previously calculated in \cite{DelDuca:2004wt,Anastasiou:2011qx,DelDuca:2015zqa,Mondini:2019vub,Mondini:2019gid} and one-loop amplitudes with up to four partons in \cite{Dixon:2009uk,Badger:2009hw,Badger:2009vh,Anastasiou:2011qx,DelDuca:2015zqa,Mondini:2019vub,Mondini:2019gid}.
The $Hgg$ two-loop three-parton amplitudes in the heavy-top limit were first derived in \cite{Gehrmann:2011aa} by some of us in the context of NNLO corrections to Higgs production in conjunction with a jet at hadron colliders \cite{Chen:2014gva,Chen:2016zka}.
The corresponding two-loop amplitudes for $H\to b\bar{b} g$ have been calculated in \cite{Ahmed:2014pka,Mondini:2019vub}.
We have verified that our calculation agrees with the numerical results of \cite{Mondini:2019vub}.
\begin{table}[t]
  \centering
  \caption{Partonic channels contributing to the decay $H \to 3j$ up to \NNLO.}
  \begin{tabular}{llll}\toprule
    ~ & $H\to b\bar{b}$ & $H\to gg$ & ~\\ \midrule
    \LO & ~ & $H \to ggg$ & tree-level \\
    ~ & $H \to b\bar{b}g$ & $H \to g q\bar{q}$ & tree-level \\ \midrule
    \NLO & ~ & $H \to g g g$ & one-loop \\
    ~ & $H \to b b g$ & $H \to  g q \bar{q}$ & one-loop \\
    ~ & ~ & $H \to g g g g$ & tree-level \\
    ~ & $H \to b \bar{b} g g$ & $H \to g g q \bar{q}$ & tree-level \\
    ~ & $H \to b \bar{b} q \bar{q}$ & $H \to q \bar{q} q' \bar{q}'$ & tree-level \\
    ~ & $H \to b \bar{b} b \bar{b}$ & $H \to q \bar{q} q \bar{q}$ & tree-level \\ \midrule
    \NNLO & ~ & $H \to g g g$ & two-loop \\
    ~ & $H \to b b g$ & $H \to g q \bar{q}$ & two-loop \\
    ~ & ~ & $H \to g g g g$ & one-loop \\
    ~ & $H \to b \bar{b} g g$ & $H \to g g q \bar{q}$ & one-loop \\
    ~ & $H \to b \bar{b} q \bar{q}$ & $H \to q \bar{q} q' \bar{q}'$ & one-loop \\
    ~ & $H \to b \bar{b} b \bar{b}$ & $H \to q \bar{q} q \bar{q}$ & one-loop \\
    ~ & ~ & $H \to g g g g g$ & tree-level \\
    ~ & $H \to b \bar{b} g g g$ & $H \to g g g q \bar{q}$ & tree-level \\
    ~ & $H \to b \bar{b} q \bar{q} g$ & $H \to g q \bar{q} q' \bar{q}'$ & tree-level \\
    ~ & $H \to b \bar{b} b \bar{b} g$ & $H \to g q \bar{q} q \bar{q}$ & tree-level \\
    \bottomrule
  \end{tabular}
  \label{tab:channels}
\end{table}

\subsection{Formalism}
\label{subsec:formalism}
We perform the computation in the rest frame of the Higgs boson and only take its decay into account.
In this configuration, the only mass-scale in the process is the Higgs boson rest mass $m_H$.  For an infrared-safe observable $y$, the differential decay rate of the Higgs boson to three-jet-like final states 
in the decay channel $X$, 
normalised to the respective Born-level $H\to X$ decay width ($X=b\bar{b}$, $gg$), can be written up to \NNLO in the strong coupling $\alphas$ as,
\begin{eqnarray}\label{eq:exp1}
  \frac{1}{\Gamma^{(0)}_{H\to X}(m_H,\mu_R)}\frac{\rd \Gamma_{H\to X}(m_H,\mu_R)}{\rd y}&=&\nonumber \\ &&\hspace{-6cm}\left(\frac{\alphas(\mu_R)}{2\uppi}\right)\frac{\rd A(\mu_R)}{\rd y} + \left(\frac{\alphas(\mu_R)}{2\uppi}\right)^2\frac{\rd B(\mu_R)}{\rd y} + \left(\frac{\alphas(\mu_R)}{2\uppi}\right)^3\frac{\rd C(\mu_R)}{\rd y} + \mathcal{O}(\alpha_s^4)\,.
  \label{eq:rateDiff}
\end{eqnarray}
where $\mu_R$ is the renormalisation scale. $A$, $B$, and $C$ denote the dimensionless \LO, \NLO, and \NNLO coefficients, respectively.
Given a suitable lower cutoff $y_0$ on the observable, the \LO coefficient $A$ is finite for $y > y_0$.
The calculation of the \NLO and \NNLO coefficients $B$ and $C$, however, requires a suitable scheme to remove explicit infrared singularities in virtual matrix elements and implicit infrared divergences in real-radiation contributions.
Here, we employ the antenna-subtraction scheme \cite{Gehrmann-DeRidder:2005btv,Currie:2013vh}, in which real radiation subtraction terms are assembled from antenna functions.
Subsequently, the real subtraction counterterms are integrated over the radiation phase space to yield virtual subtraction terms, which cancel the explicit poles in the virtual corrections.
Specifically, we construct antenna-subtraction terms from generalised antenna functions \cite{Fox:2024bfp}, which are in turn derived  directly from the relevant infrared limits using the algorithm described in \cite{Braun-White:2023sgd,Braun-White:2023zwd}.

At \NLO and \NNLO, a consistent prediction is obtained from~\eqref{eq:rateDiff} upon normalising by the inclusive \NLO or \NNLO decay width $\Gamma^{(1)}_{H\to X}$ or $\Gamma^{(2)}_{H\to X}$, respectively.
Writing the Higgs-boson decay rates in terms of their \LO results~\eqref{eq:ratesLO} as
\begin{align}
  \Gamma^{(k)}_{H\to b\bar{b}} &= \Gamma^{(0)}_{H\to b\bar{b}}\, \left(1+\sum\limits_{n=1}^k\alphas^n C_{b\bar{b}}^{(n)}\right)\,,\\
  \Gamma^{(k)}_{H\to gg} &= \Gamma^{(0)}_{H\to gg}\, \left(1+\sum\limits_{n=1}^k\alphas^n C_{gg}^{(n)}\right)\,,
  \label{eq:ratesNkLO}
\end{align}
and expanding the normalisation we obtain at NNLO
\begin{eqnarray}\label{eq:exp2}
	\frac{1}{\Gamma^{(2)}_{H\to X}(m_H,\mu_R)}\frac{\rd \Gamma_{H\to X}(m_H,\mu_R)}{\rd y}&=&\nonumber \\ &&\hspace{-6cm}\left(\frac{\alphas(\mu_R)}{2\uppi}\right)\frac{\rd \bar{A}(\mu_R)}{\rd y} + \left(\frac{\alphas(\mu_R)}{2\uppi}\right)^2\frac{\rd \bar{B}(\mu_R)}{\rd y} + \left(\frac{\alphas(\mu_R)}{2\uppi}\right)^3\frac{\rd \bar{C}(\mu_R)}{\rd y} + \mathcal{O}(\alpha_s^4) \,.
	\label{eq:rateDiffNorm}
\end{eqnarray}
The coefficients $\bar{A}$, $\bar{B}$, and $\bar{C}$ are given in terms of Eq.~\eqref{eq:ratesNkLO} by
\begin{equation}
  \bar{A} = A\,, \quad \bar{B} = B - C^{(1)}A\,, \quad \bar{C} = C - C^{(1)}B + \left(\left(C^{(1)}\right)^2-C^{(2)}\right)A \,.
\end{equation}
The relevant higher-order corrections to the inclusive decay widths are for example given in \cite{Herzog:2017dtz} and we report them explicitly in Appendix~\ref{app:A}. For completeness we also illustrate the renormalisation scale dependence of the expansion coefficients in~\eqref{eq:exp1} and~\eqref{eq:exp2} in Appendix~\ref{app:B}.

In Section~\ref{subsec:predictions}, we present results for the quantity $y\frac{1}{\Gamma_{H\to X}^{(k)}(m_H,\mu_R)}\frac{\rd \Gamma(m_H,\mu_R)}{\rd y}$, where the weighting by a factor $y$ with respect to~\eqref{eq:exp2}  damps the differential distributions for small values of a generic event shape $y$. In event-shape distribution studies in $Z$-boson decays~\cite{Jones:2003yv}, a central scale $\mu_R=m_Z$ 
is chosen as default. Inclusive Higgs boson production through gluon fusion at the LHC is conventionally evaluated~\cite{LHCHiggsCrossSectionWorkingGroup:2016ypw} at central 
renormalisation and factorisation scales $\mu_R=\mu_F=m_H/2$. This difference in conventions reflects the generic 
ambiguity in setting scales in fixed-order perturbation theory. It is motivated by the larger size of 
QCD corrections in Higgs boson processes, in particular from its gluonic coupling mode. 
We will employ $\mu_R=m_H/2$ as central scale setting in Section~\ref{sec:results} below, and estimate the 
uncertainty on the prediction at a given perturbative order by varying $\mu_R$ by a factor 2 around its central value. 

\subsection{Observables}
We consider distributions in  the following event-shape variables in our study.

\paragraph{Three-jet resolution}
In a given jet algorithm, the three-jet resolution variable determines the value of the respective distance measure at which the event is reclustered from a three-jet event to a two-jet event, $y_{23} := \min_{(i,j)}\{y_{ij}\}$ among all pairs $(i,j)$ in a three-jet configuration.
In the Durham jet algorithm \cite{Catani:1991hj,Brown:1990nm,Brown:1991hx,Stirling:1991ds,Bethke:1991wk}, the particle-distance measure is defined as
\begin{equation}
  y_{ij}^{\mathrm{D}} = \frac{2\min (E_i^2,E_j^2)(1-\cos\theta_{ij})}{s} \, .
\end{equation}
Here, we consider the ``E-scheme'', in which the four-momentum of a pseudo-jet consisting of a pair $i$, $j$ is calculated as the sum of four-momenta $p_i$ and $p_j$.

\paragraph{$C$-Parameter}
The three eigenvalues $\lambda_1$, $\lambda_2$, and $\lambda_3$ of the linearised momentum tensor \cite{Parisi:1978eg,Donoghue:1979vi},
\begin{equation}
  \Theta^{\alpha\beta} = \frac{1}{\sum\limits_j\mods{\vec{p}_j}}\sum\limits_i\frac{p_i^\alpha p_i^\beta}{\mods{\vec{p}_i}} \, ,\;\; \text{where } \alpha,\beta \in \{1,2,3\} \,,
\end{equation}
define the $C$-parameter,
\begin{equation}
  C = 3(\lambda_1\lambda_2 + \lambda_2\lambda_3 + \lambda_3\lambda_1) \,.
\end{equation}
The $\lambda_i$ are defined such that $0\leq \lambda_{1,2,3}\leq 1$, $\sum_i\lambda_i=1$ and therefore $0\leq C\leq 1$.

\paragraph{Thrust}
In multi-particle final states, the thrust axis $\vec{n}_T$ is defined as the unit vector which maximises the thrust definition \cite{Brandt:1964sa,Farhi:1977sg},
\begin{equation}
  T = \max\limits_{\vec{n}}\left(\frac{\sum\limits_i \mods{\vec{p}_i\cdot\vec{n}}}{\sum\limits_i \mods{\vec{p}_i}} \right) \,.
\end{equation}
The thrust observable is directly related to the degree of isotropy of collider events.
The thrust approaches unity, $T\to 1$, in the two-particle limit and is bounded from below by $T\geq\frac{1}{2}$, which it approaches in the fully isotropical multi-particle limit.
For events containing exactly three particles, it holds that $T \geq \frac{2}{3}$.
In order to mirror the limiting behaviour of the other event-shape variables, which 
all vanish in the two-jet limit, 
$T$ is typically replaced by 
\begin{equation}
  \tau \equiv 1-T = \min\limits_{\vec{n}}\left(1-\frac{\sum\limits_i \mods{\vec{p}_i\cdot\vec{n}}}{\sum\limits_i \mods{\vec{p}_i}} \right) \, ,
\end{equation}
so that $\tau > 0$ corresponds to the departure from the two-particle limit and $\tau \leq \frac{1}{2}$.

\paragraph{Heavy jet mass}
The heavy jet mass makes use of the two hemispheres $\mathcal{H}_\mathrm{L}$ and $\mathcal{H}_\mathrm{R}$, defined by the plane orthogonal to the thrust axis.
It is defined as \cite{Clavelli:1981yh}
\begin{equation}
  \rho_\mathrm{H} \equiv \frac{M_\mathrm{H}^2}{s} = \max_{i\in\{\mathrm{L},\mathrm{R}\}}\left(\frac{M_i^2}{s}\right) \,,
\end{equation}
where the scaled invariant hemisphere masses $M_{\mathrm{L}/\mathrm{R}}$ are given by
\begin{equation}
  \frac{M_{\mathrm{L}/\mathrm{R}}^2}{s} = \frac{1}{s}\left(\sum\limits_{j \in \mathcal{H}_{\mathrm{L}/\mathrm{R}}} p_j\right)^2 \, .
\end{equation}
For three-parton final states, the definition of $\rho_\mathrm{H}$ coincides with $\tau$, thus yielding identical distributions at \LO. 

\paragraph{Jet broadenings}
In terms of the momenta residing in the two hemispheres $\mathcal{H}_{\mathrm{L}/\mathrm{R}}$ separated by the plane orthogonal to the thrust axis $\vec{n}_T$, the hemisphere broadenings can be defined as
\begin{equation}
  B_{\mathrm{L}/\mathrm{R}} = \frac{\sum\limits_{j\in \mathcal{H}_{\mathrm{L}/\mathrm{R}}}\mods{\vec{p}_j\times \vec{n}_T}}{2\sum\limits_j\mods{\vec{p}_j}} \,.
\end{equation}
From these, two different three-jet-like event-shape observables, the total and wide jet broadening $B_\mathrm{T}$ and $B_\mathrm{W}$, can be defined \cite{Rakow:1981qn,Catani:1992jc}
\begin{equation}
  B_\mathrm{T} = B_\mathrm{L}+B_\mathrm{R} \,, \quad B_\mathrm{W} = \max(B_\mathrm{L},B_\mathrm{R}) \,.
\end{equation}
The two-particle limit is characterised by vanishing jet broadenings, $B_\mathrm{T}\to 0$, \mbox{$B_\mathrm{W} \to 0$}.
Both obersables are bounded from above by $B_\mathrm{T} = B_\mathrm{W} = \frac{1}{2\sqrt{3}}\approx 0.2887$ for events containing exactly three particles.

\paragraph{Soft-drop thrust}
In the context of lepton colliders, the soft-drop algorithm \cite{Larkoski:2014wba} is separately applied to the two hemispheres $\mathcal{H}_{\mathrm{L}}$ and $\mathcal{H}_{\mathrm{R}}$.
For each of the two hemispheres, the Cambridge--Aachen algorithm \cite{Dokshitzer:1997in,Wobisch:1998wt} is applied until exactly one (hemisphere) jet is left.
Subsequently, the angular-ordered clustering sequence of the Cambridge-Aachen algorithm is recursively declustered, during which soft pseudo-jets are groomed from the resulting hemisphere jet.
In each hemisphere, the soft-drop algorithm performs the following steps:
\begin{enumerate}
\item split the pseudo-jet corresponding to the last step of the clustering algorithm into its constituents $i$ and $j$;
\item test the soft-drop criterion
  \begin{equation}
    \frac{\min\left(E_i,E_j\right)}{E_i+E_j} > z_\mathrm{cut} \theta_{ij}^\beta \, ,
  \end{equation}
  where $z_\mathrm{cut}$ and $\beta \geq 0$ denote the soft-drop parameters, $\theta_{ij}$ is the angle between the pseudo-jets $i$ and $j$, and $E_i$, $E_j$ are their energies;
\item if the constituents $i$ and $j$ fail the soft-drop criterion, discard (groom) the softer of the two and restart the algorithm for the remaining (harder) pseudo-jet;
\item if the constituents $i$ and $j$ pass the soft-drop criterion, terminate the algorithm; the resulting hemisphere jet is the combination of the pseudo-jets $i$ and $j$.
\end{enumerate}
In terms of the soft-drop groomed hemispheres $\mathcal{H}_{\mathrm{L}/\mathrm{R}}^\mathrm{SD}$, the soft-drop groomed event $\mathcal{E}_\mathrm{SD}$, and the ungroomed event $\mathcal{E}$, the soft-drop thrust is defined as \cite{Baron:2018nfz,Marzani:2019evv}
\begin{equation}
  \tau_\mathrm{SD} = \frac{\sum\limits_{i\in \mathcal{E}_\mathrm{SD}}\mods{\vec p_i}}{\sum\limits_{i\in \mathcal{E}}\mods{\vec p_i}}\left(1 - \frac{\sum\limits_{i \in \mathcal{H}^\mathrm{SD}_\mathrm{L}}\mods{\vec n_\mathrm{L} \cdot \vec p_i} + \sum\limits_{i \in \mathcal{H}^\mathrm{SD}_\mathrm{R}}\mods{\vec n_\mathrm{R} \cdot \vec p_i}}{\sum\limits_{i\in\mathcal{E}_\mathrm{SD}}\mods{\vec p_i}} \right) \,.
\end{equation}

\section{Results}
\label{sec:results}
In this section, we present the main results of our calculation.
We summarise our numerical setup and parameter choices in Sec.~\ref{subsec:setup}, before presenting event-shape distributions in the $H\to b\bar{b}$ and $H\to gg$ channels up to NNLO in Sec.~\ref{subsec:predictions}.

\subsection{Numerical Setup}
\label{subsec:setup} 
Electroweak parameters are considered constant and set in the $G_\upmu$ scheme, with input values
\begin{equation}
  \GF =  1.1664\times 10^{-5}~\GeV^{-2}\,, \quad m_Z = 91.200~\GeV \,, % \,, \quad m_W = 80.379~\GeV \,.
\end{equation}
where $\GF$ determines the (effective) $Hgg$ and $Hb\bar{b}$ couplings $\lambda_0$ and $y_b$ via the vacuum expectation value $v=({\sqrt{2}\GF})^{-\tfrac{1}{2}}=246.22$ GeV, and $m_Z$ serves as the reference scale for the strong coupling and all quark masses in the $\overline{\text{MS}}$ scheme.
As discussed above, the central scale is chosen as $\muR = m_H/2 = 125.09/2~\GeV$ and perturbative uncertainties are assessed by varying this scale as $\muR \to k_\mu \muR$ with $k_\mu \in \left[\frac{1}{2},2\right]$. 
The strong coupling $\alphas$ is evaluated at one, two, or three loops at \LO, \NLO, and \NNLO, respectively, with a nominal value at the $Z$-boson mass of $\alphas(m_Z) = 0.11800$.

We present differential distributions for $y\frac{1}{\Gamma_{H\to X}^{(k)}(m_H,\mu_R)}\frac{\rd \Gamma(m_H,\mu_R)}{\rd y}$ (see~\eqref{eq:exp2}). For each event shape we show the results for the Yukawa-induced decay to bottom quarks, the decay to gluons and the total sum of all decay modes. For the gluonic decay mode, we rescale $\lambda_0^2(\mu_R)$ in~\eqref{eq:couplings} to include finite top, bottom and charm mass effects in the effective coupling~\cite{Spira:1997dg}, as well as electroweak corrections~\cite{Actis:2008ug}. In the sum, we also include the decay to charm quarks, which is formally identical to the decay to bottom quarks after rescaling by a factor $y_c(\mu_R)^2/y_b(\mu_R)^2$, but has a significant impact on the phenomenological predictions.  The total sum over decay modes is then defined at a given perturbative order $k$ as:
\begin{equation}
	y\frac{1}{\Gamma^{(k)}(m_H,\mu_R)}\sum_{X}\frac{\rd \Gamma(m_H,\mu_R)_{H\to X}}{\rd y},\quad\text{with}\,X=b\bar{b},c\bar{c},gg,
\end{equation}
where the total hadronic decay width is given by
\begin{equation}
	\Gamma^{(k)}=\Gamma^{(k)}_{H\to b\bar{b}}+\Gamma^{(k)}_{H\to c\bar{c}}+\Gamma^{(k)}_{H\to gg}.
\end{equation}
We work with running bottom and charm quark Yukawa couplings in the  $\overline{\text{MS}}$ scheme  \mbox{$y_b(m_H)=m_b(m_H)/v=0.011309$}, \mbox{$y_c(m_H)=m_c(m_H)/v=0.0024629$} and $\overline{\text{MS}}$ top quark mass \mbox{$m_t(m_H) = 166.48~\GeV$}. The quark masses and Yukawa couplings are evolved to different scales following~\cite{Vermaseren:1997fq}.

The event-shape distributions are ill-defined in fixed-order perturbation theory for $y\to 0$, where large 
logarithmic corrections must be resummed to all orders in the coupling constant. We only display the 
distributions up to the point where the perturbative approach breaks down.
This is identified by the value of $y$ for which the differential distributions in the gluonic channel turn negative, which  typically happens earlier than in the Yukawa channel, as observed in~\cite{Fox:2025cuz} for the jet rates in hadronic Higgs decays. 

\begin{figure}[htbp]
	\centering
	\includegraphics[width=0.45\linewidth]{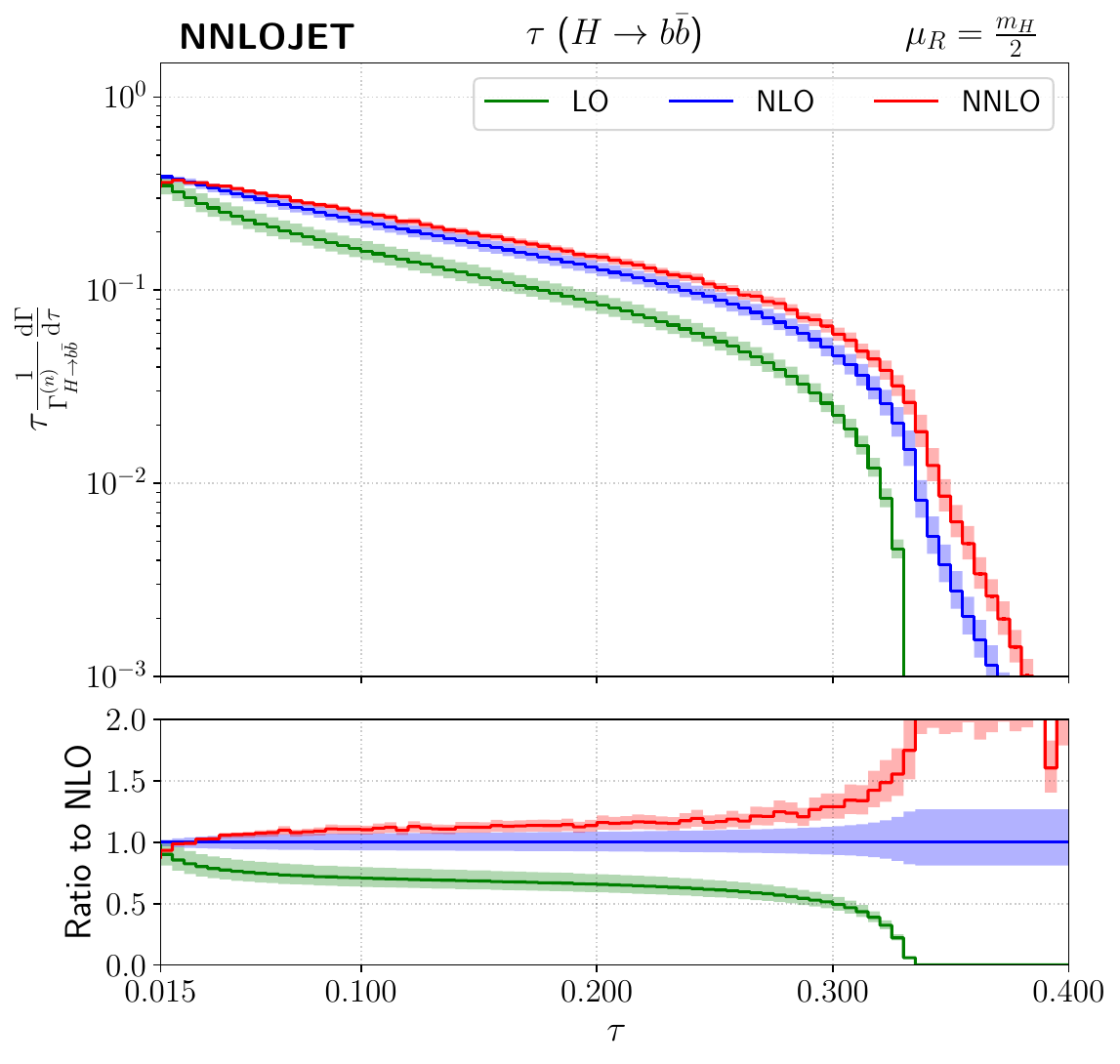}
	\includegraphics[width=0.45\textwidth]{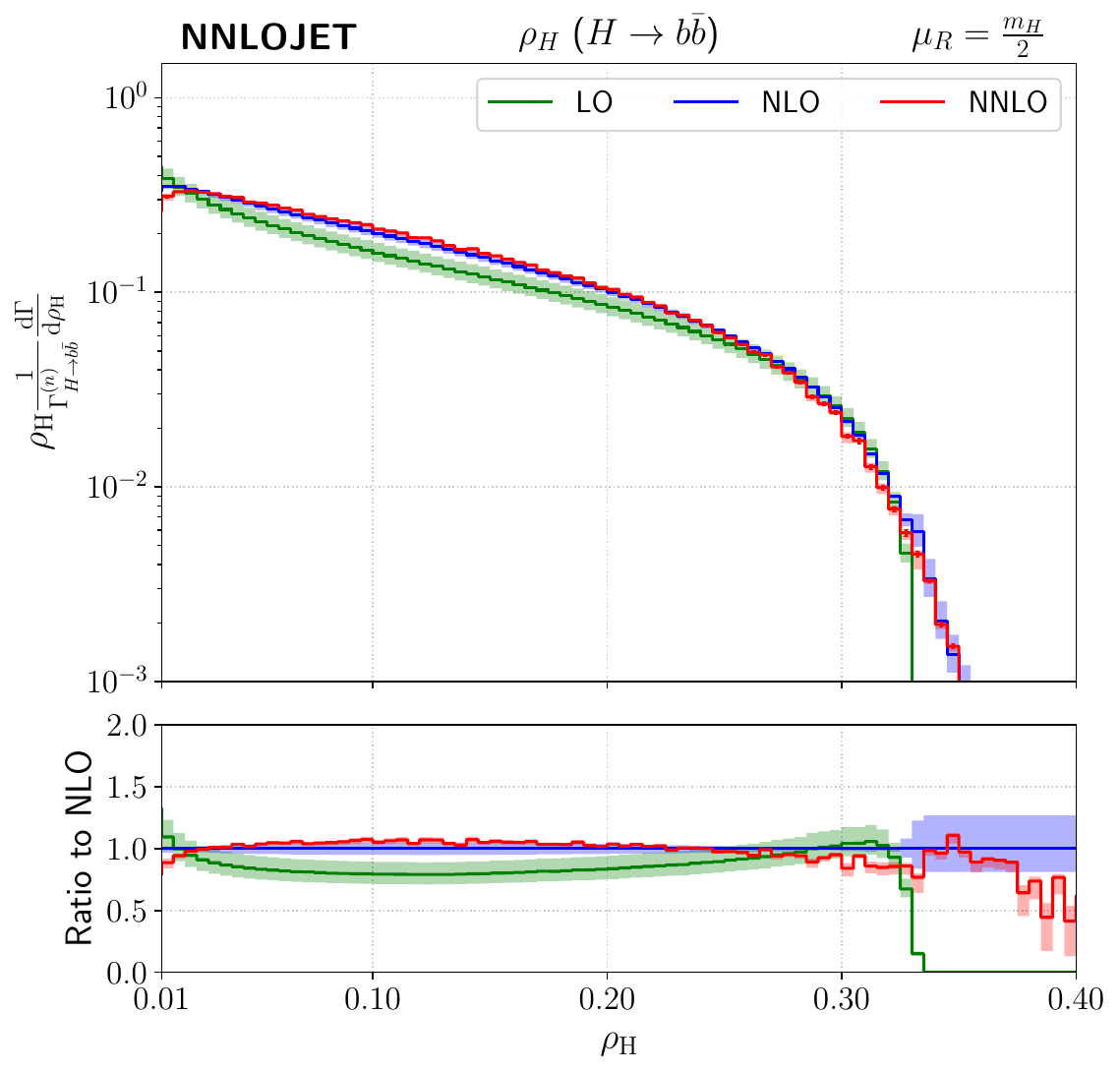}\\
	\includegraphics[width=0.45\linewidth]{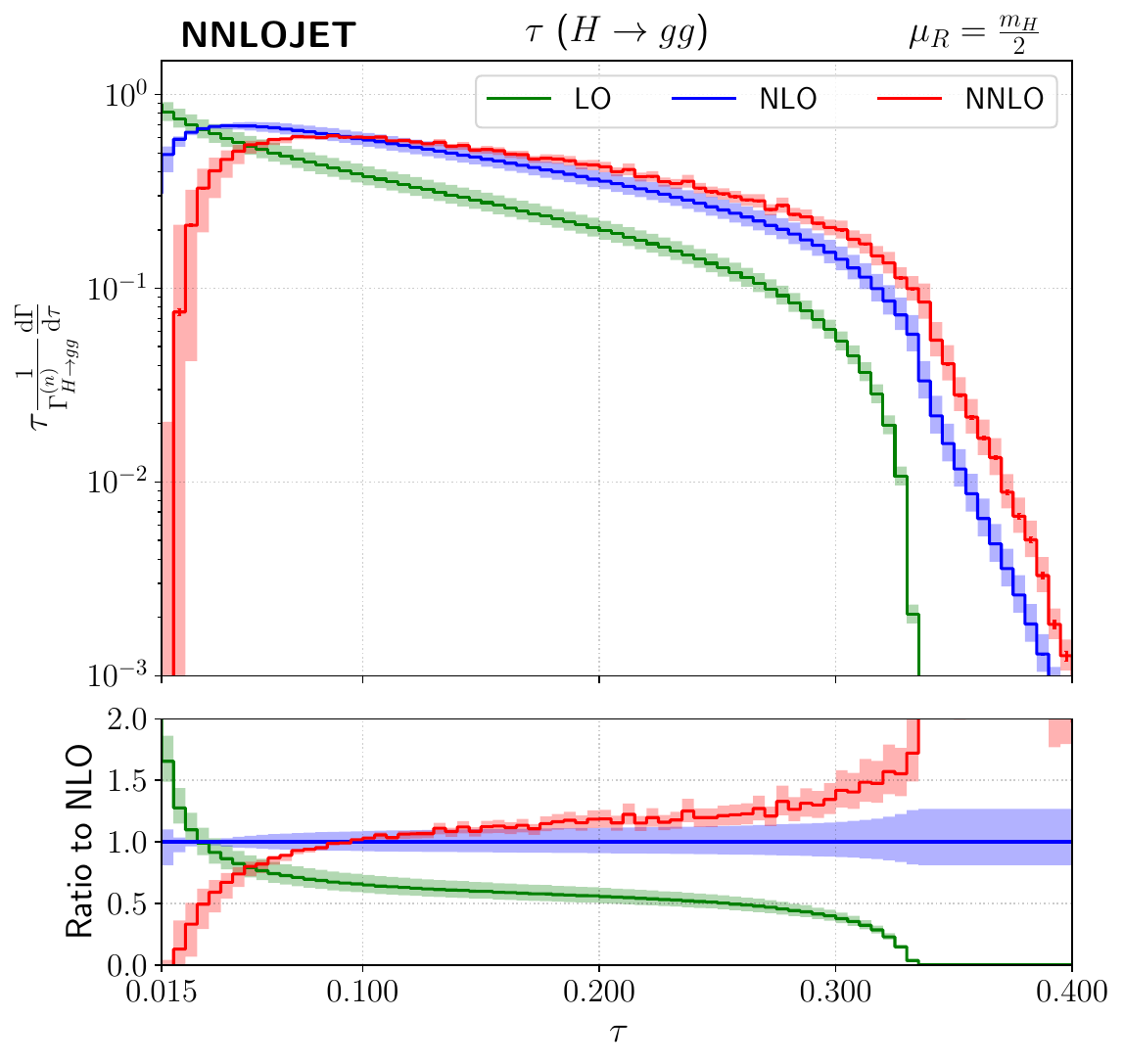}
	\includegraphics[width=0.45\textwidth]{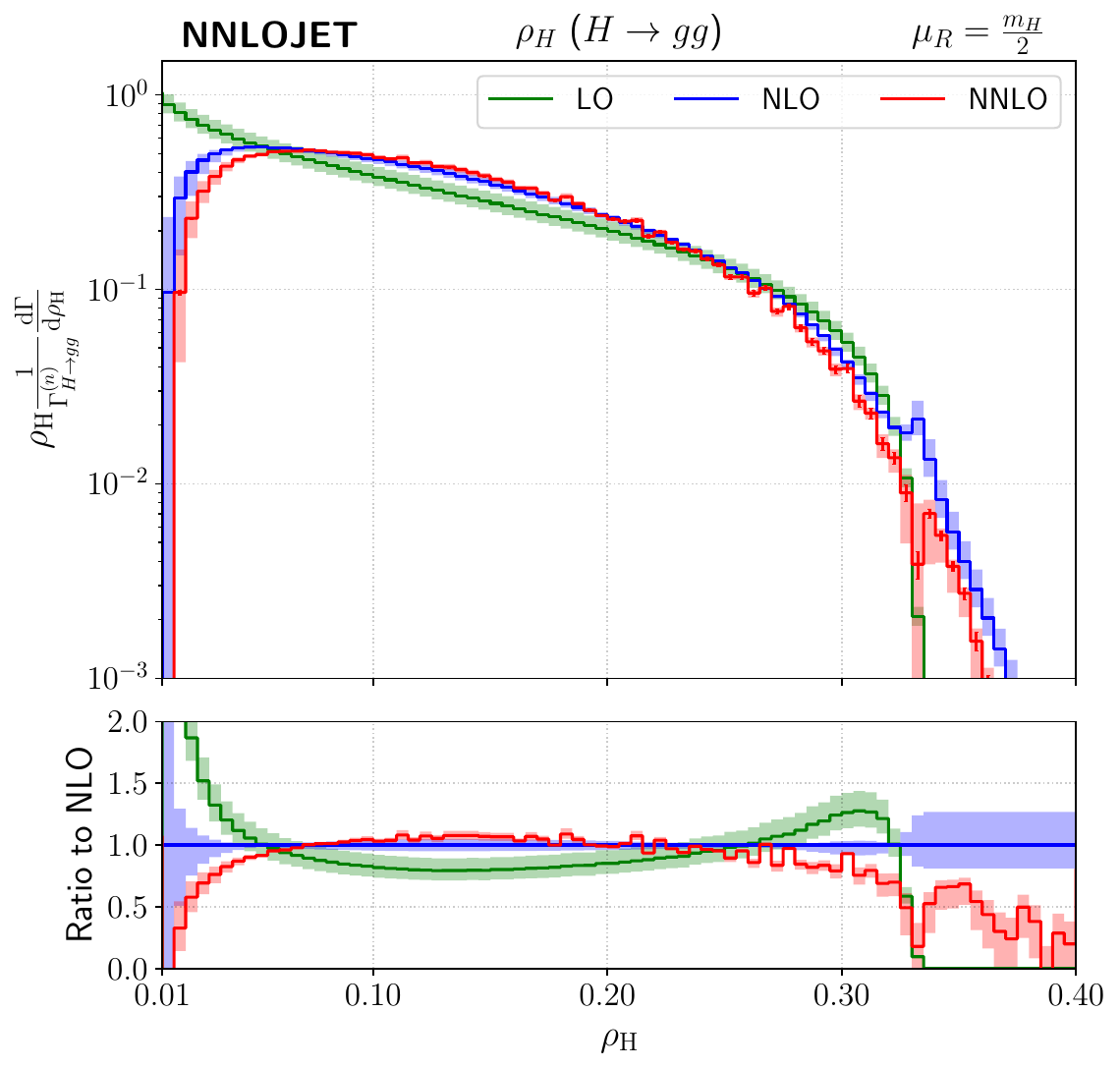}\\
	\includegraphics[width=0.45\linewidth]{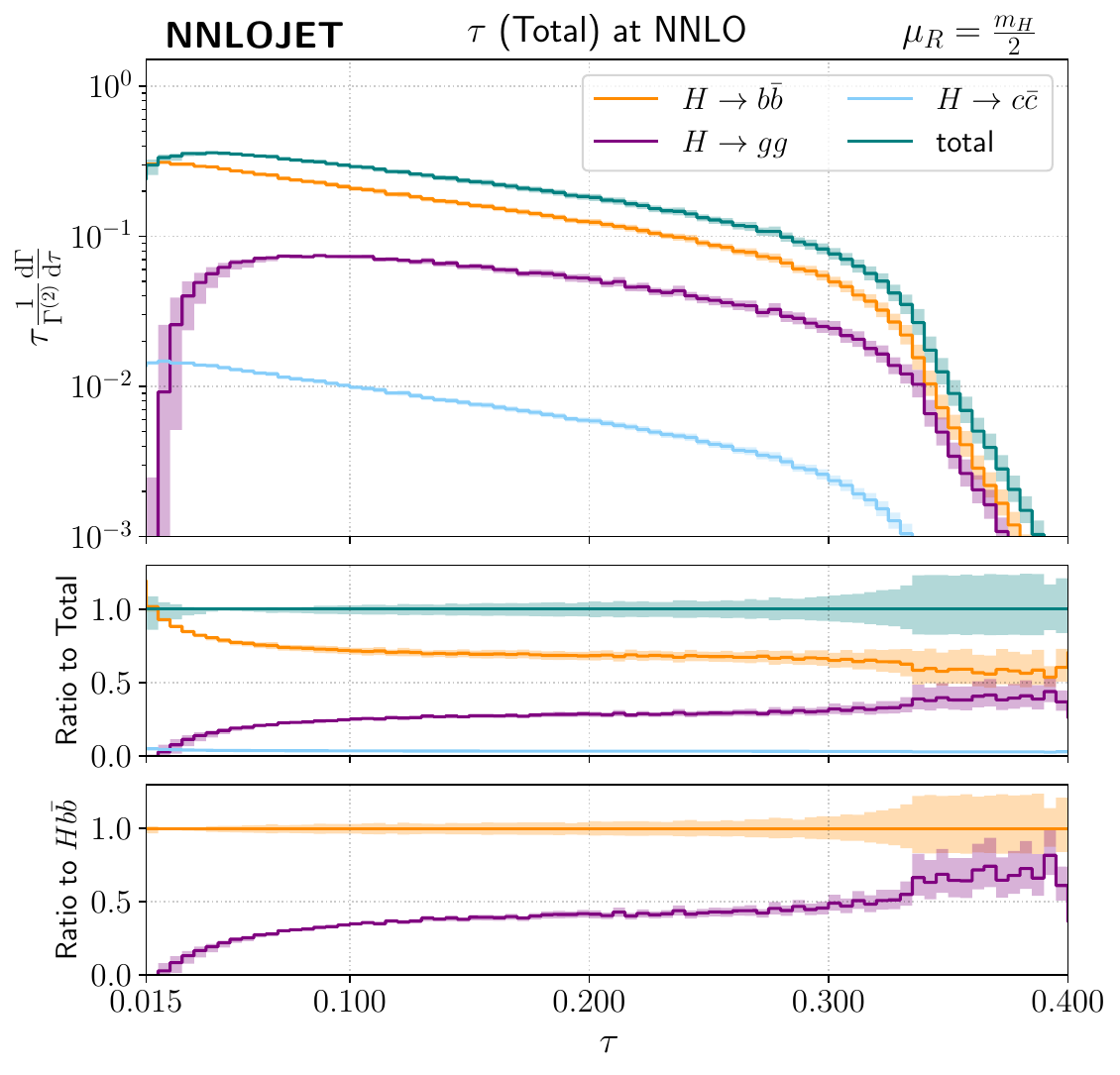}
	\includegraphics[width=0.45\textwidth]{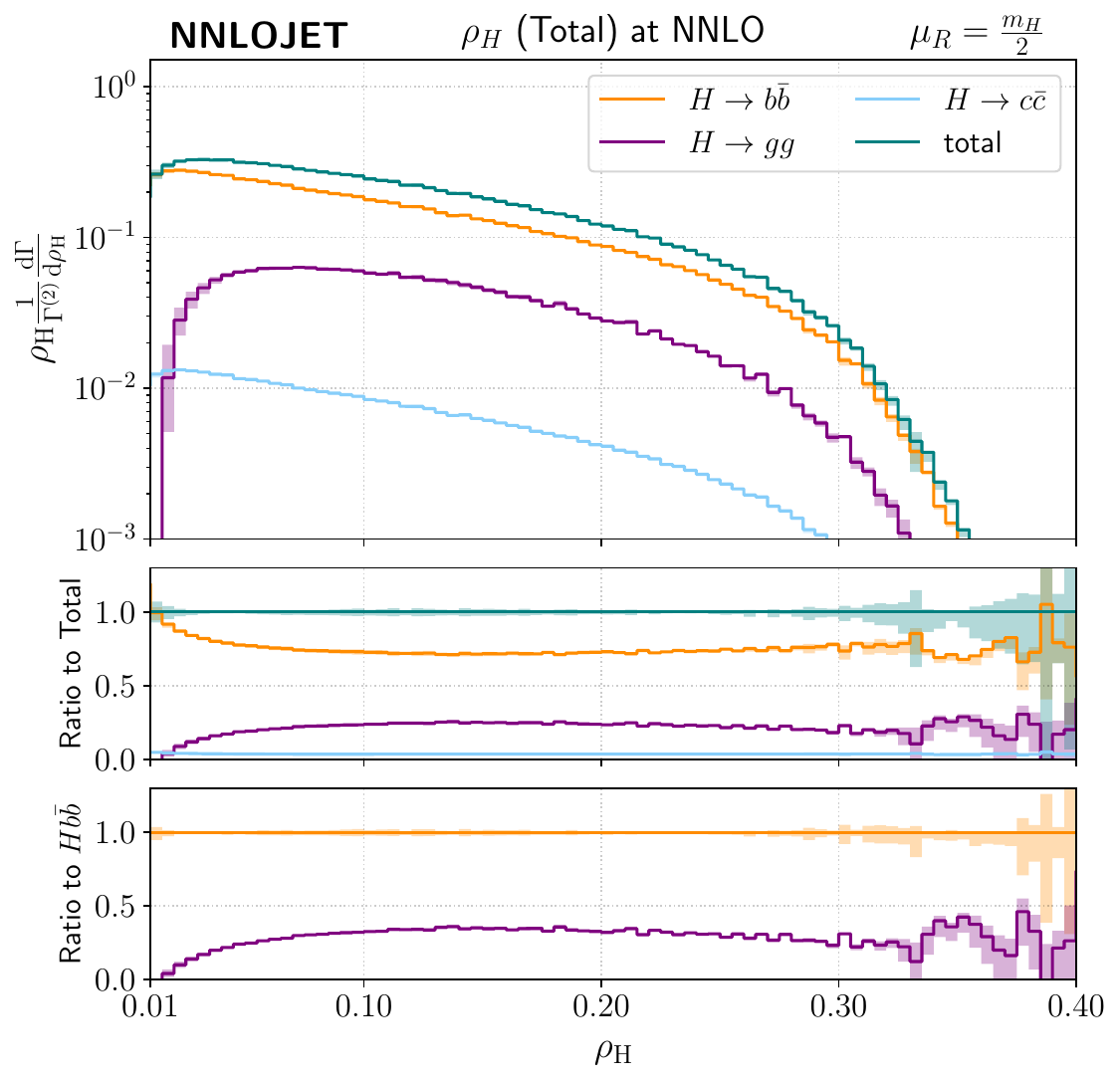}
	\caption{Thrust (left column) and heavy jet mass (right column) distributions for the $H\to b\bar{b}$ channel (top row), the $H\to gg$ channel (middle row) and the sum of all decay modes at NNLO (bottom row). In the top and middle row, the LO (green), NLO (blue), and NNLO (red) are shown. In the bottom row, the individual contributions of the $H\to b\bar{b}$ (orange), $H\to gg$ (purple), and the $H\to c\bar{c}$ (light blue) channels are displayed alongside the sum (teal).}
	\label{fig:tau_HJM}
\end{figure}

\subsection{NNLO QCD predictions}
\label{subsec:predictions}

In Figure~\ref{fig:tau_HJM}, we present the results for the $\tau=1-T$ event shape and for the heavy jet mass $\rho_H$. The distributions are truncated respectively at $\tau=0.015$ and $\rho_H=0.01$, below which the gluonic mode contributions turns negative. In the bulk region of the thrust distribution $0.05<\tau<0.2$, the NNLO correction is moderate in both the Yukawa and gluonic mode, yielding an increase of less than $20\%$ with respect to the NLO result in this region. For the heavy jet mass distribution, NNLO corrections in the bulk region $0.05<\rho_H<0.25$ are smaller, typically amounting to less than $10\%$. Both distributions have a Sudakov shoulder~\cite{Catani:1997xc} at their three-parton endpoint $\tau=\rho_H=0.33$. The size of the NNLO corrections gradually increases approaching the Sudakov shoulder and in its vicinity, their behaviour changes substantially in shape and normalisation. The renormalisation-scale uncertainties decrease at NNLO, where they typically amount to $\pm(2\ldots 7)\%$  for $\tau$ and to $\pm(1\ldots 4)\%$ in $\rho_H$, with the gluonic mode typically exhibiting larger uncertainties than the Yukawa one. Perturbative convergence is good in the $\rho_H$ distribution, but poor in the $\tau$ distribution, with the NNLO and NLO uncertainty bands merely overlapping. In the range $(\tau,\rho_H)\in[0.1,0.3]$, where the relative size of the Yukawa and gluonic decay modes are roughly constant, their contribution is about $70\%$ and $30\%$ of the total for $\tau$ and $75\%$ and $25\%$ for $\rho_H$.  For large $\tau>0.3$, the gluonic mode increases to more than a third of the total, while remaining at its bulk value for $\rho_H>0.3$. In the infrared regions, $\tau\to 0$ or  $\rho_H\to 0$, the relative size of the gluonic contribution decreases, and the total is completely dominated by the Yukawa mode.

\begin{figure}[htbp]
  \centering
  \includegraphics[width=0.45\textwidth]{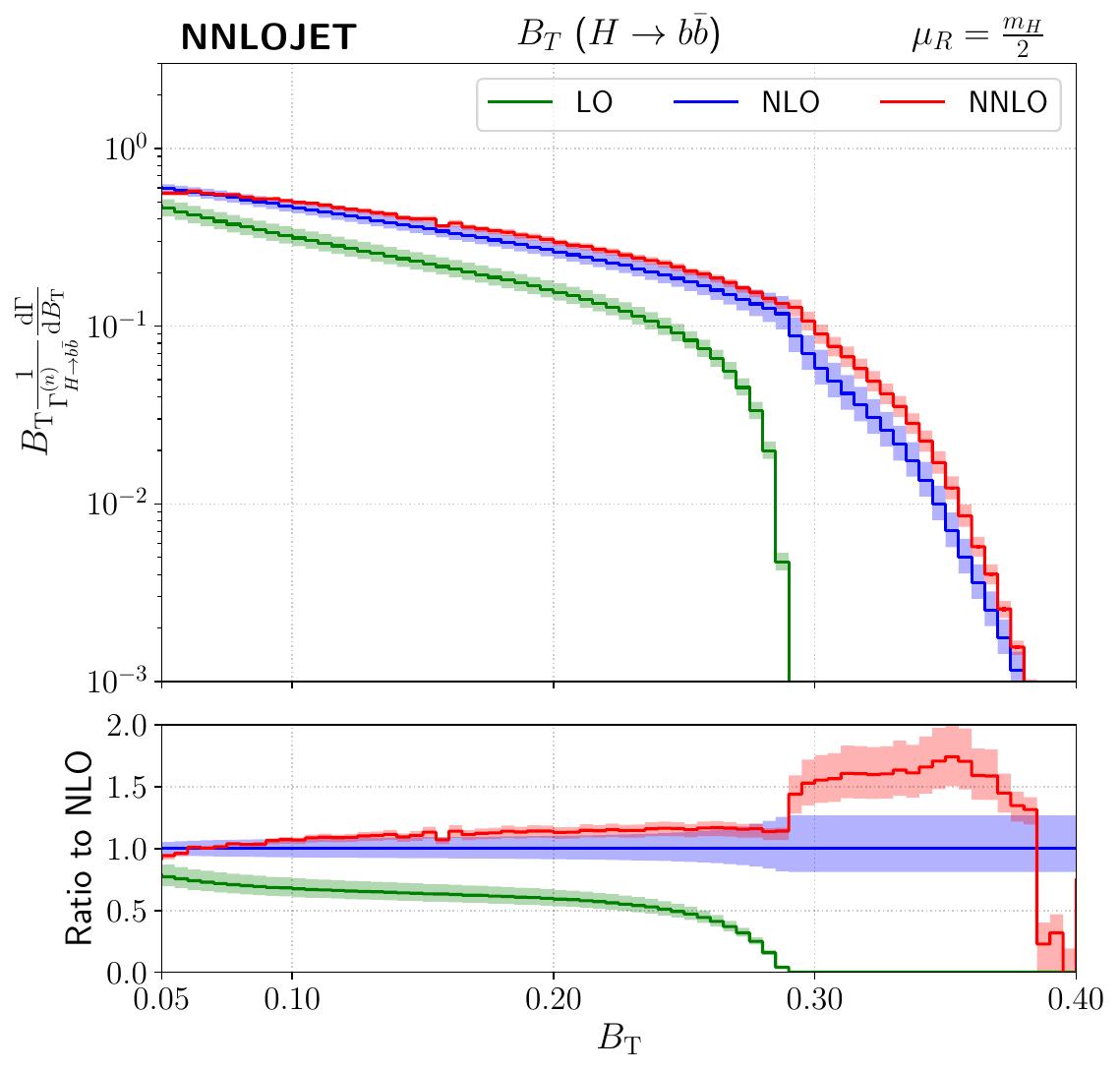}
  \includegraphics[width=0.45\textwidth]{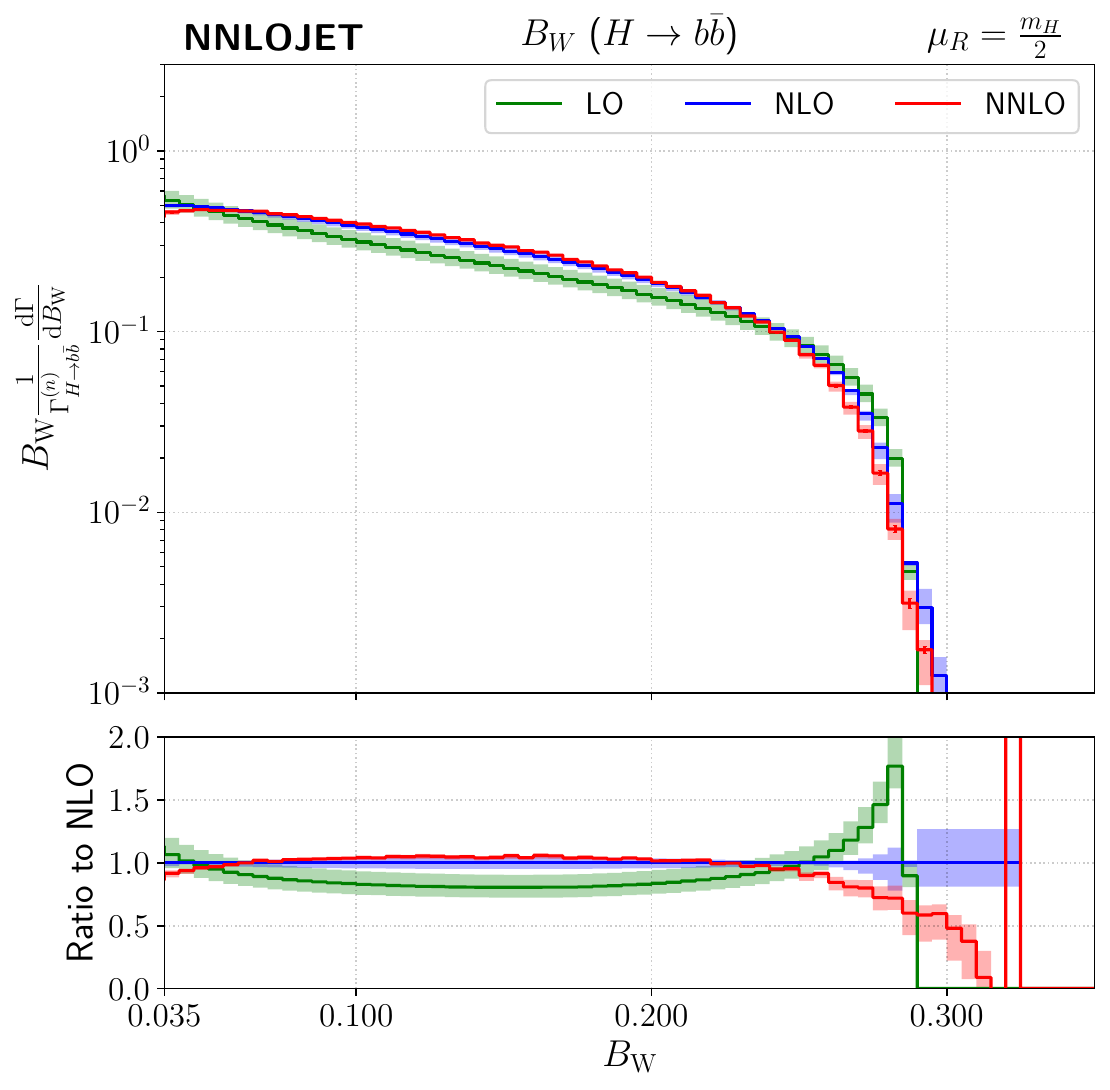}
  \includegraphics[width=0.45\textwidth]{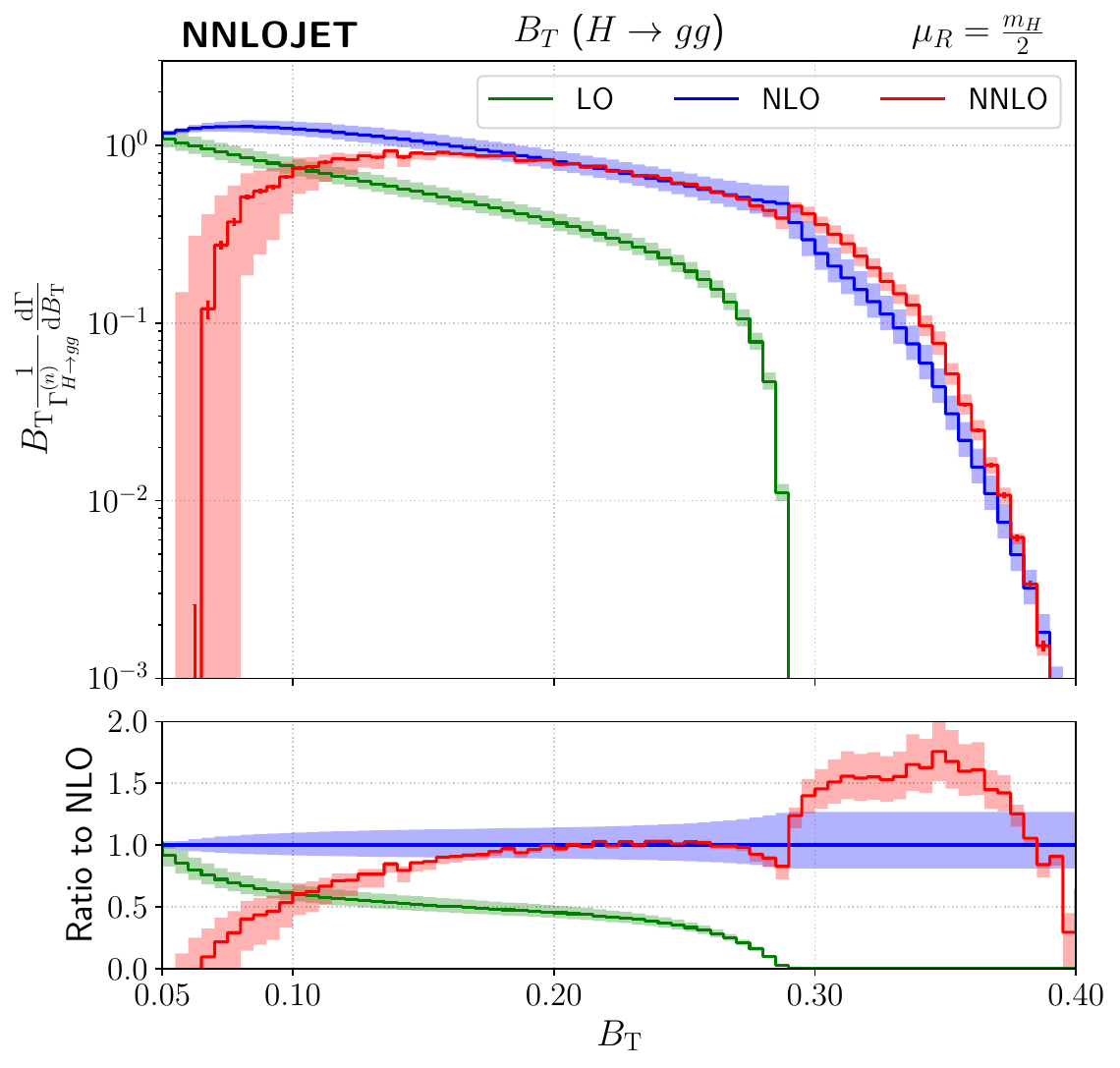}
  \includegraphics[width=0.45\textwidth]{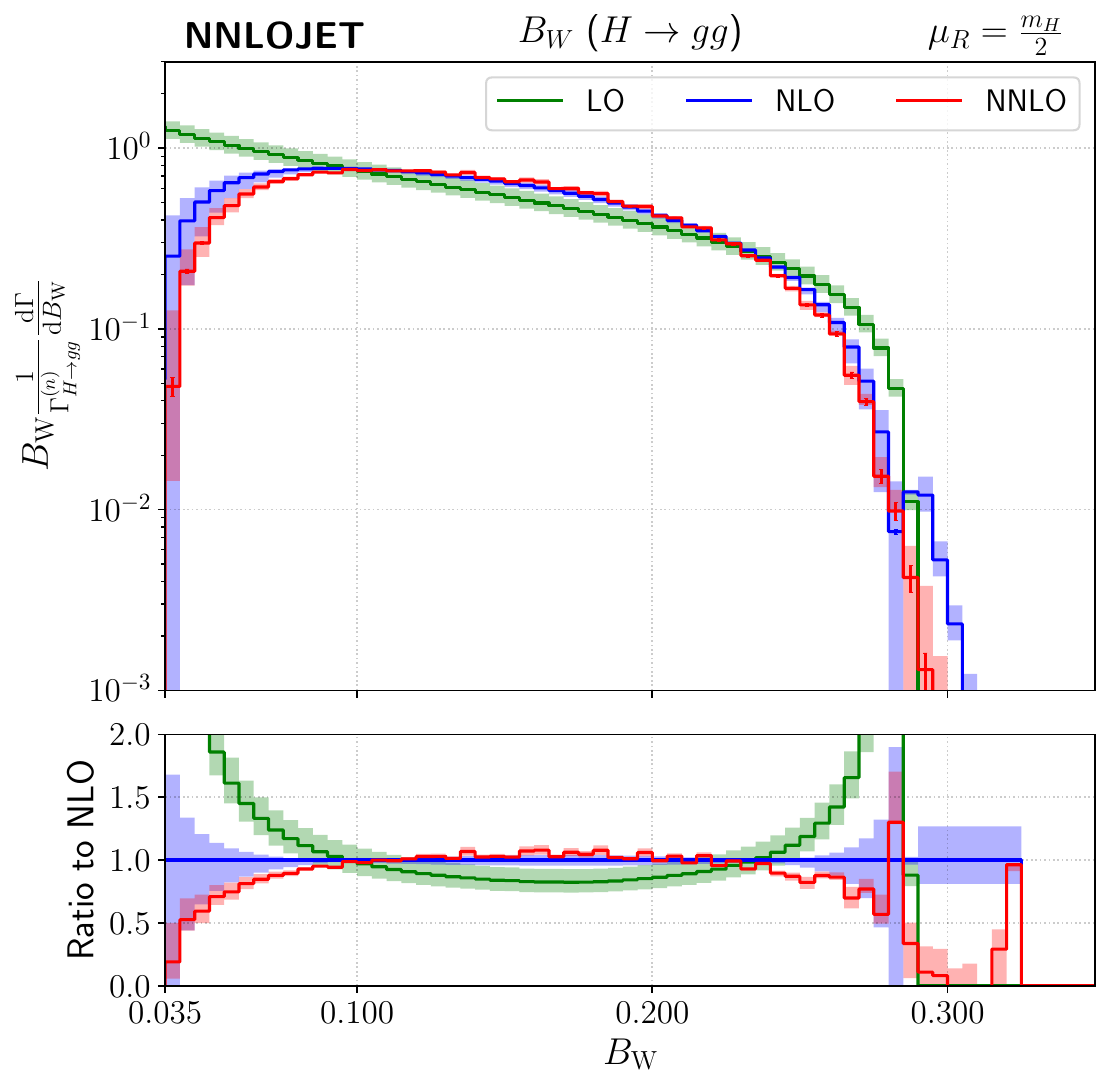}
  \includegraphics[width=0.45\textwidth]{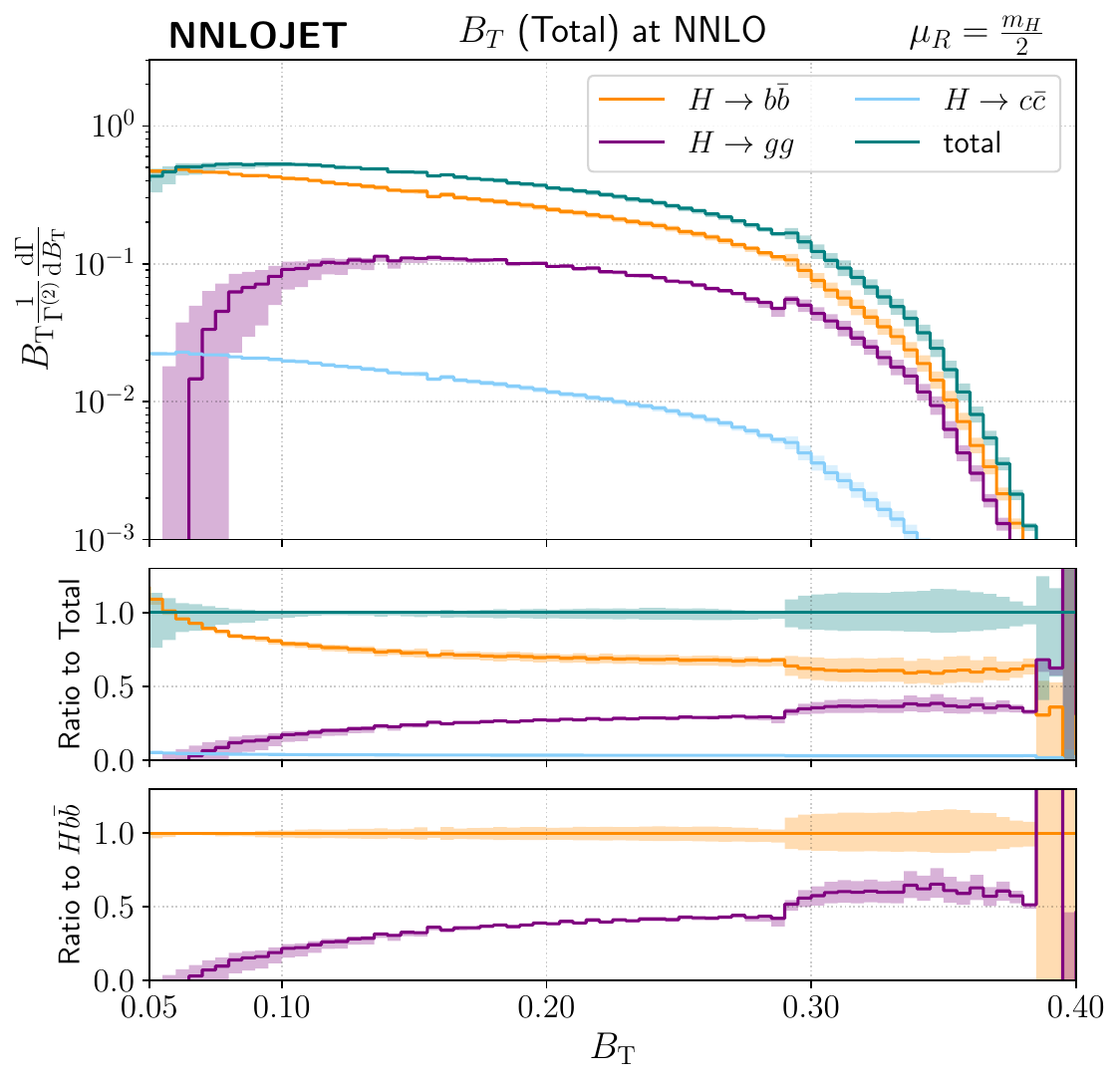}
  \includegraphics[width=0.45\textwidth]{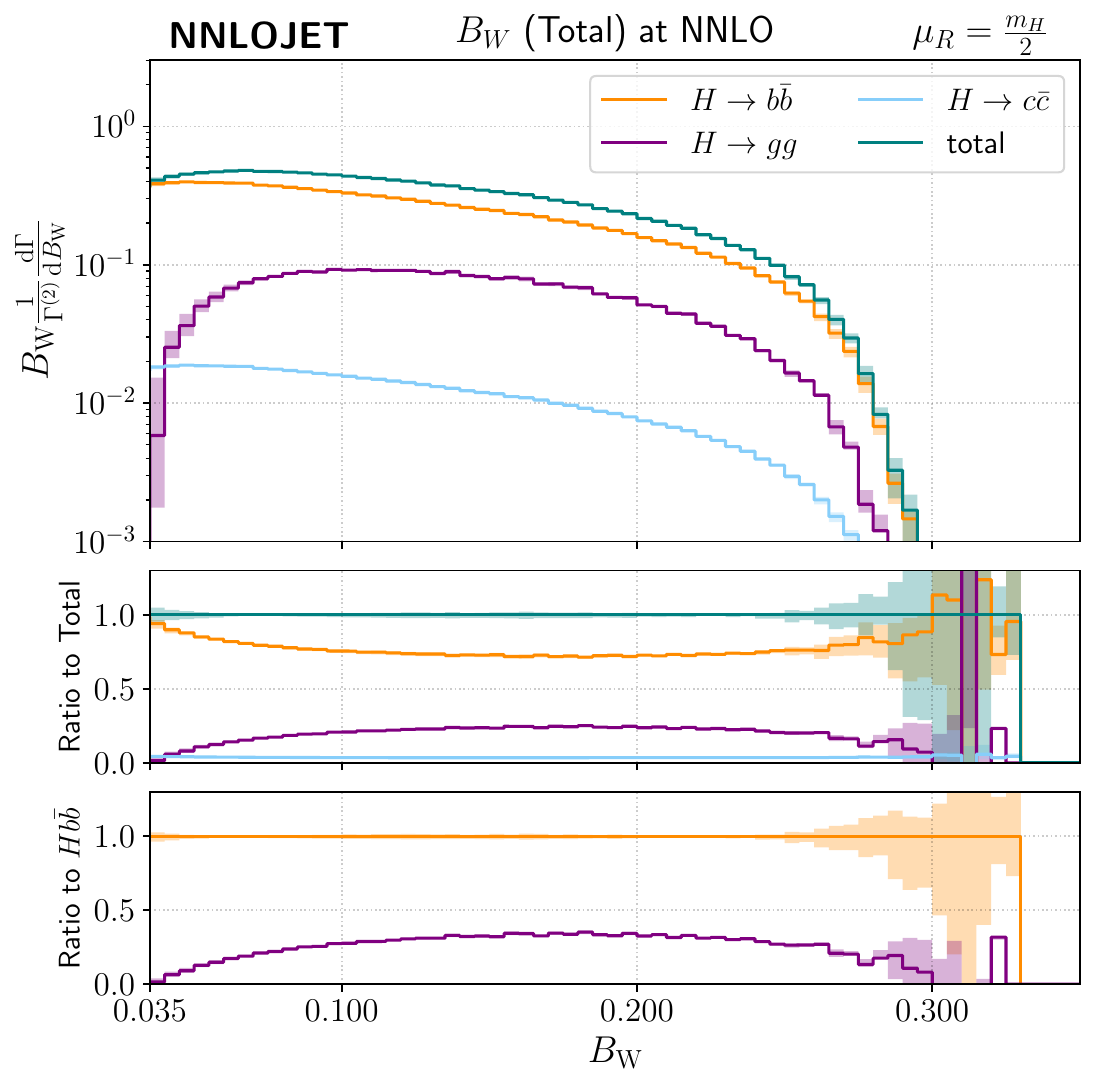}
  \caption{Total jet broadening (left column) and wide jet broadening (right column) distributions for the $H\to b\bar{b}$ channel (top row), the $H\to gg$ channel (middle row) and the sum of all decay modes at NNLO (bottom row). In the top and middle row, the LO (green), NLO (blue), and NNLO (red) are shown. In the bottom row, the individual contributions of the $H\to b\bar{b}$ (yellow), $H\to gg$ (purple) and the $H\to c\bar{c}$ (light blue) channels are displayed alongside the sum (teal).}
  \label{fig:TJB_WJB}
\end{figure}

In Figure~\ref{fig:TJB_WJB}, we present the results for the total and wide jet broadening $B_T$ and $B_W$. The distributions are truncated respectively at $B_T=0.05$ and $B_W=0.035$, below which the gluonic mode contributions turns negative. 
Both distributions have a Sudakov shoulder at $B_T=B_W=0.2887$. 
In the Yukawa mode, the bulk regions extend to  $0.05<B_{T,W}<0.25$, with mild NNLO QCD corrections 
up to $15\%$ for $B_T$ and small ones below $6\%$ for $B_W$. Perturbative convergence from NLO and NNLO is 
very good with overlapping uncertainty bands and a residual scale uncertainty of $\pm(1\dots 5)\%$ for $B_T$ and below $\pm2\%$ for $B_W$ at NNLO. 
The bulk region is smaller in the gluonic mode: $0.15<B_T<0.25$ and $0.08<B_W<0.25$, with small
NNLO corrections below $7\%$, good perturbative convergence and residual uncertainty at NNLO of $\pm(2\dots8)\%$ for $B_T$ and below $\pm4\%$ for $B_W$. Below the bulk regions, 
NNLO corrections turn negative in both modes, with an earlier onset in the gluonic mode. 
Above the Sudakov shoulder, the NNLO corrections to the 
$B_T$ distribution are large and positive, while they are large and negative in the $B_W$ distribution. The relative size of the Yukawa and gluonic channels in both the $B_T$ and $B_W$ distributions closely resembles the observations in the $\rho_H$ distribution, with contributions of about $75\%$ and $25\%$ of the total in the bulk region and 
above, and a rapid decrease of the gluonic fraction in the infrared region below.

\begin{figure}[htbp]
  \centering
  \includegraphics[width=0.45\textwidth]{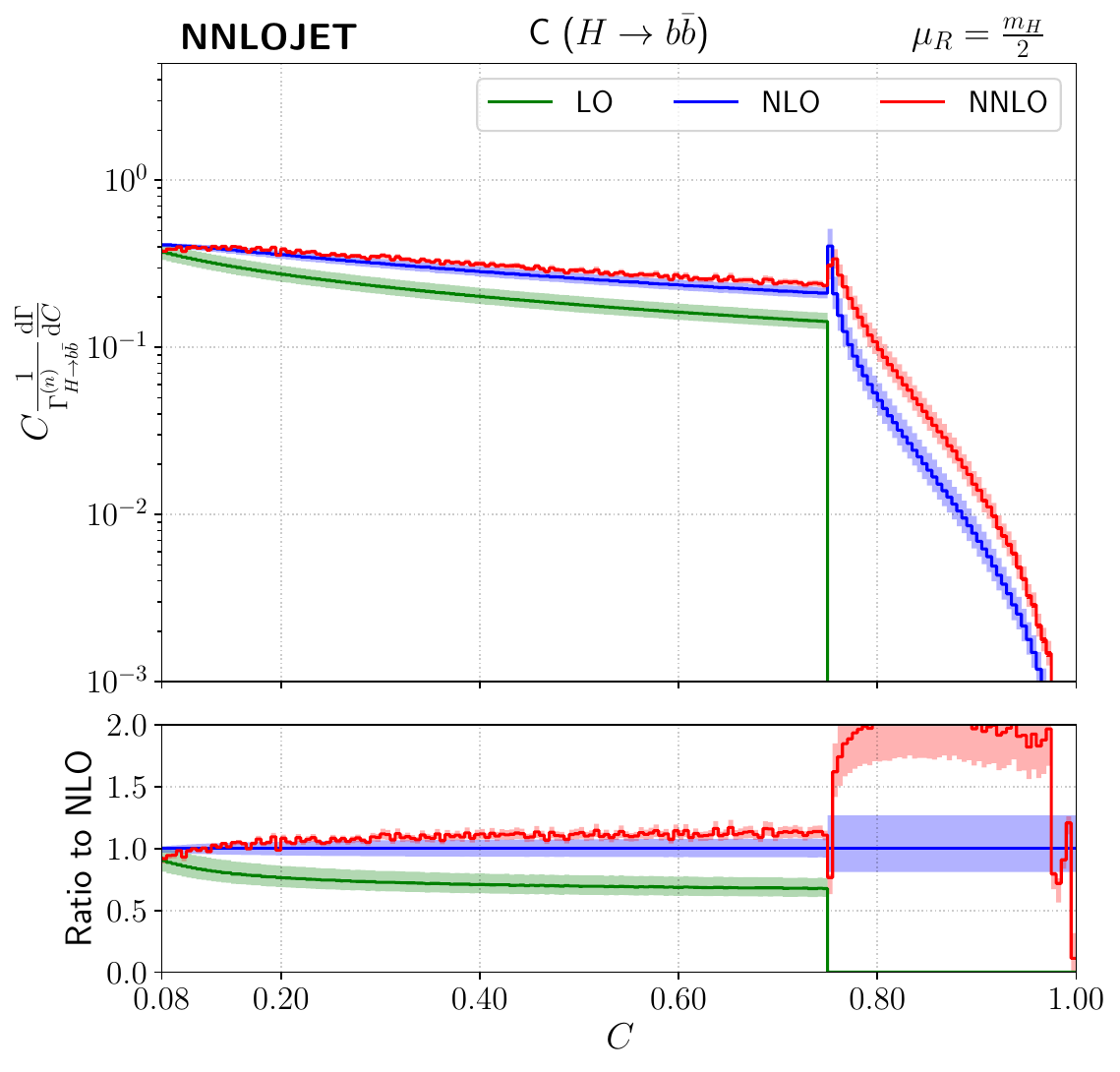}
  \includegraphics[width=0.45\textwidth]{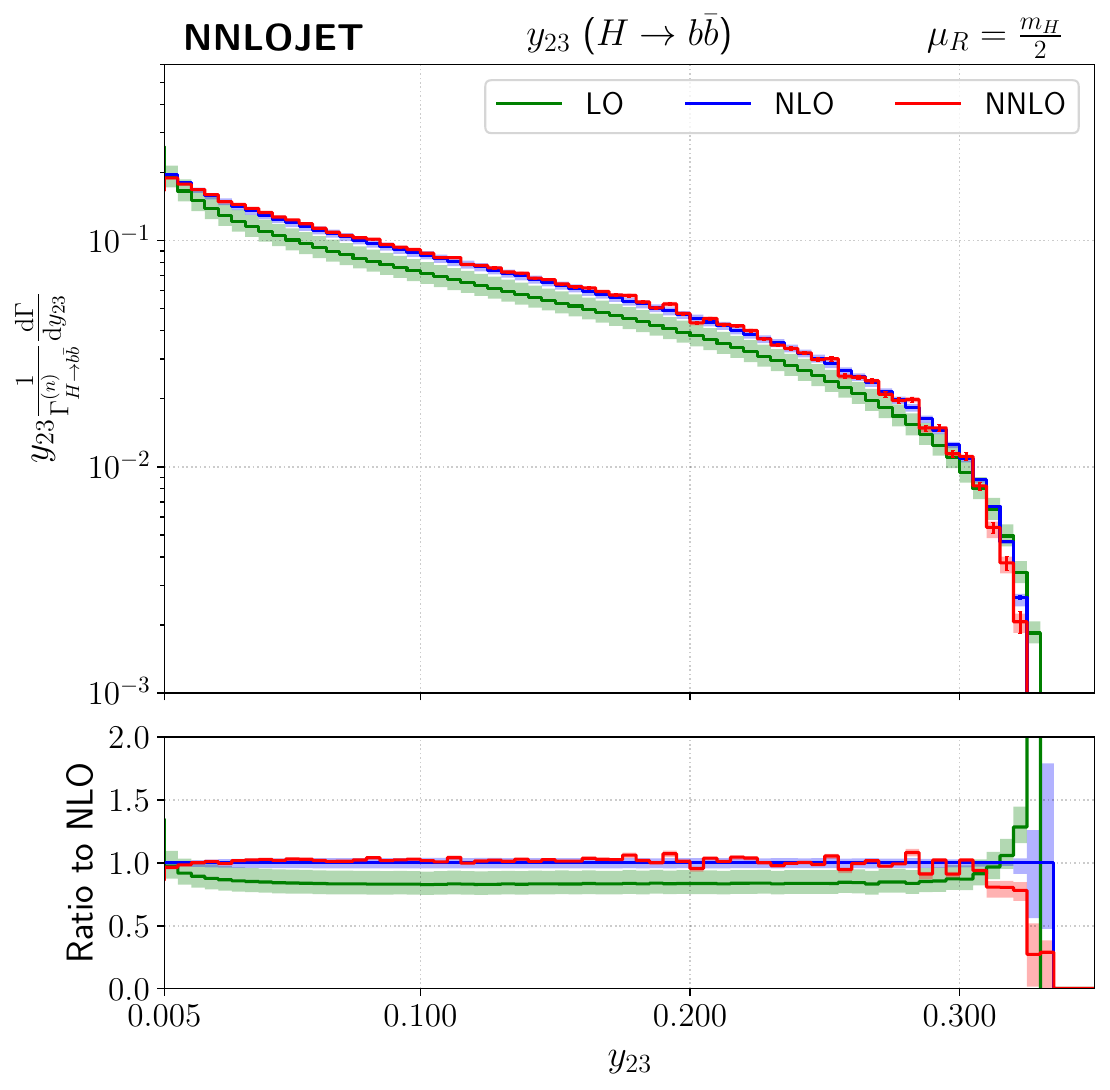}
  \includegraphics[width=0.45\textwidth]{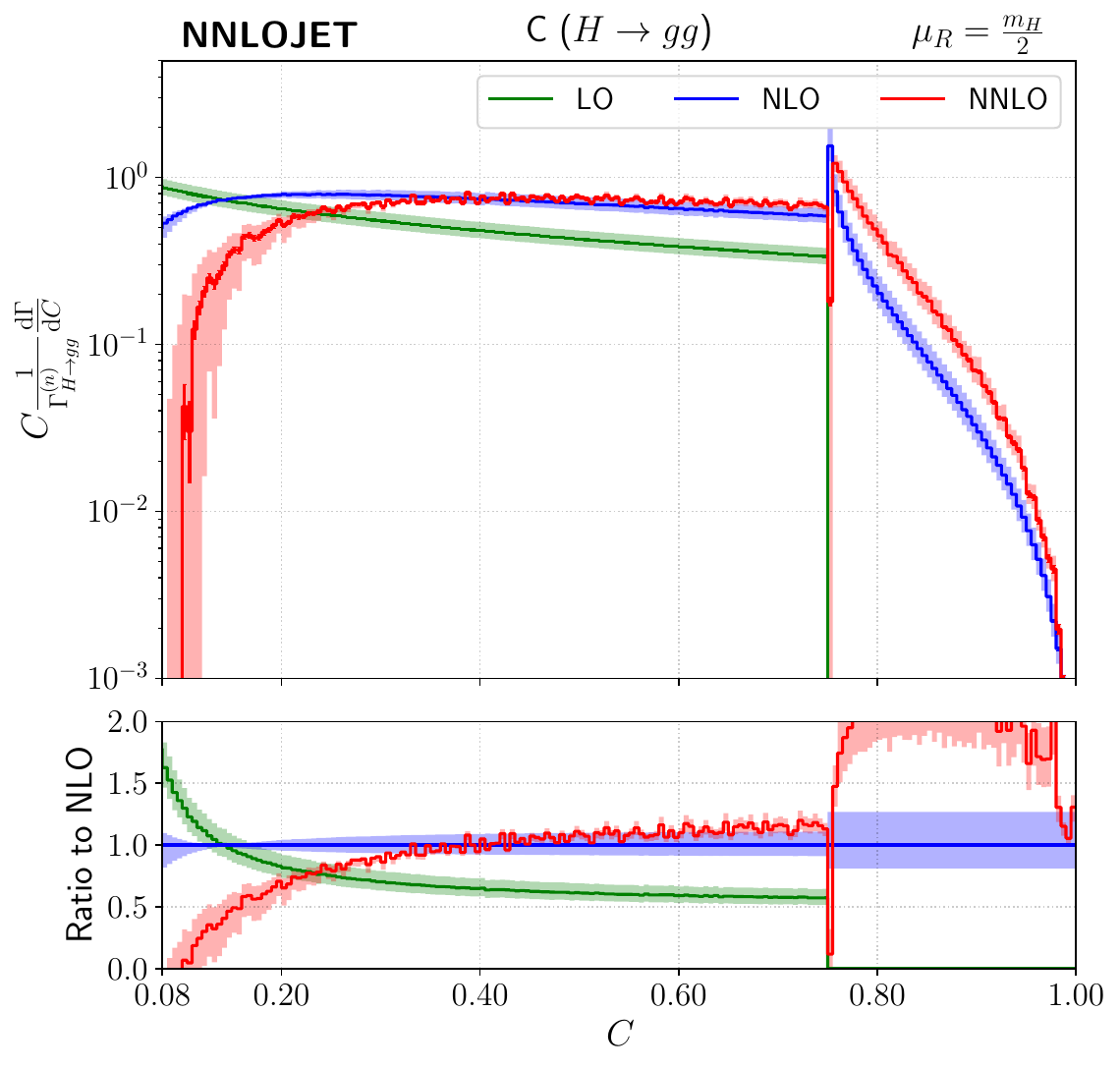}
  \includegraphics[width=0.45\textwidth]{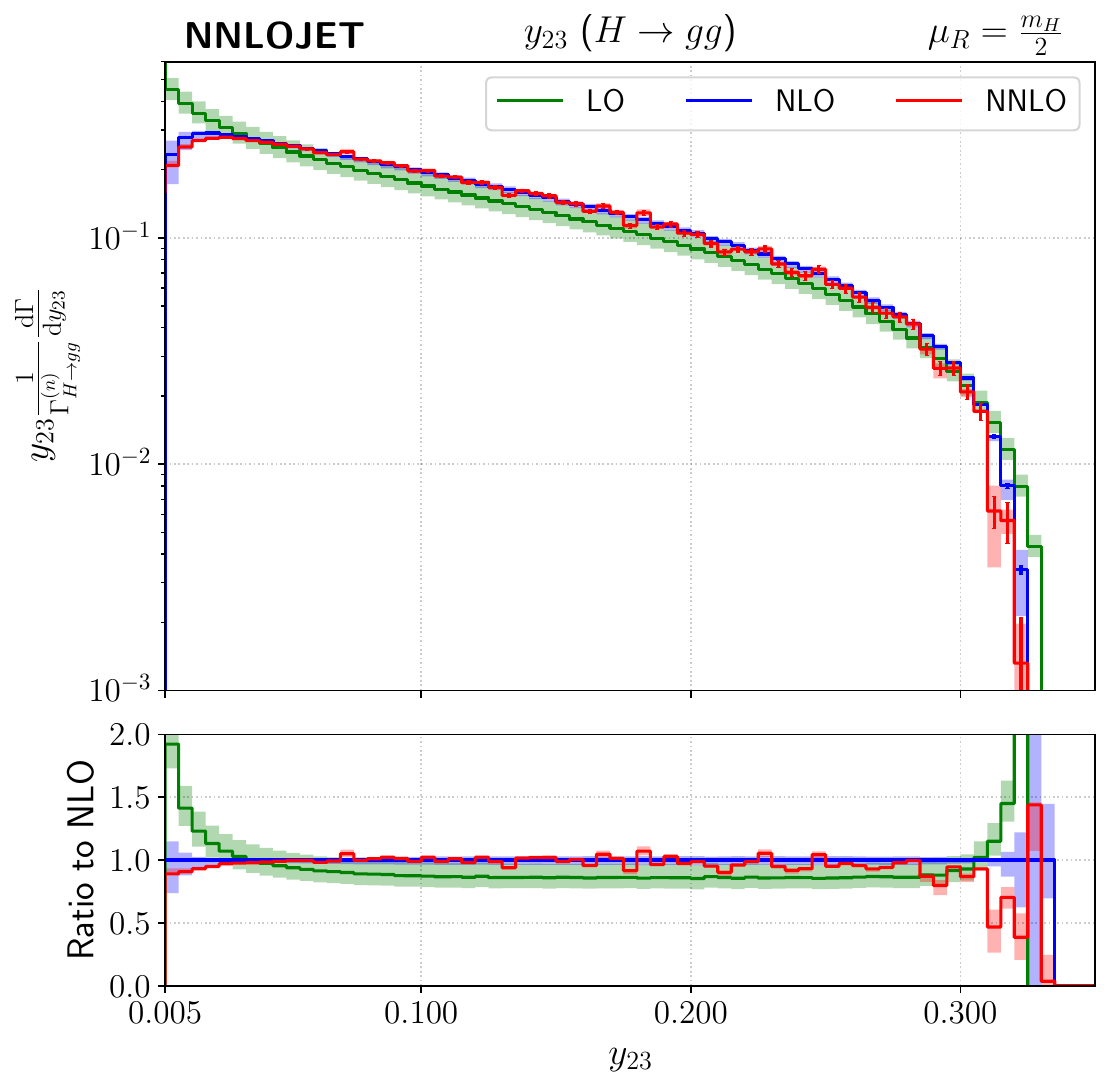}
  \includegraphics[width=0.45\textwidth]{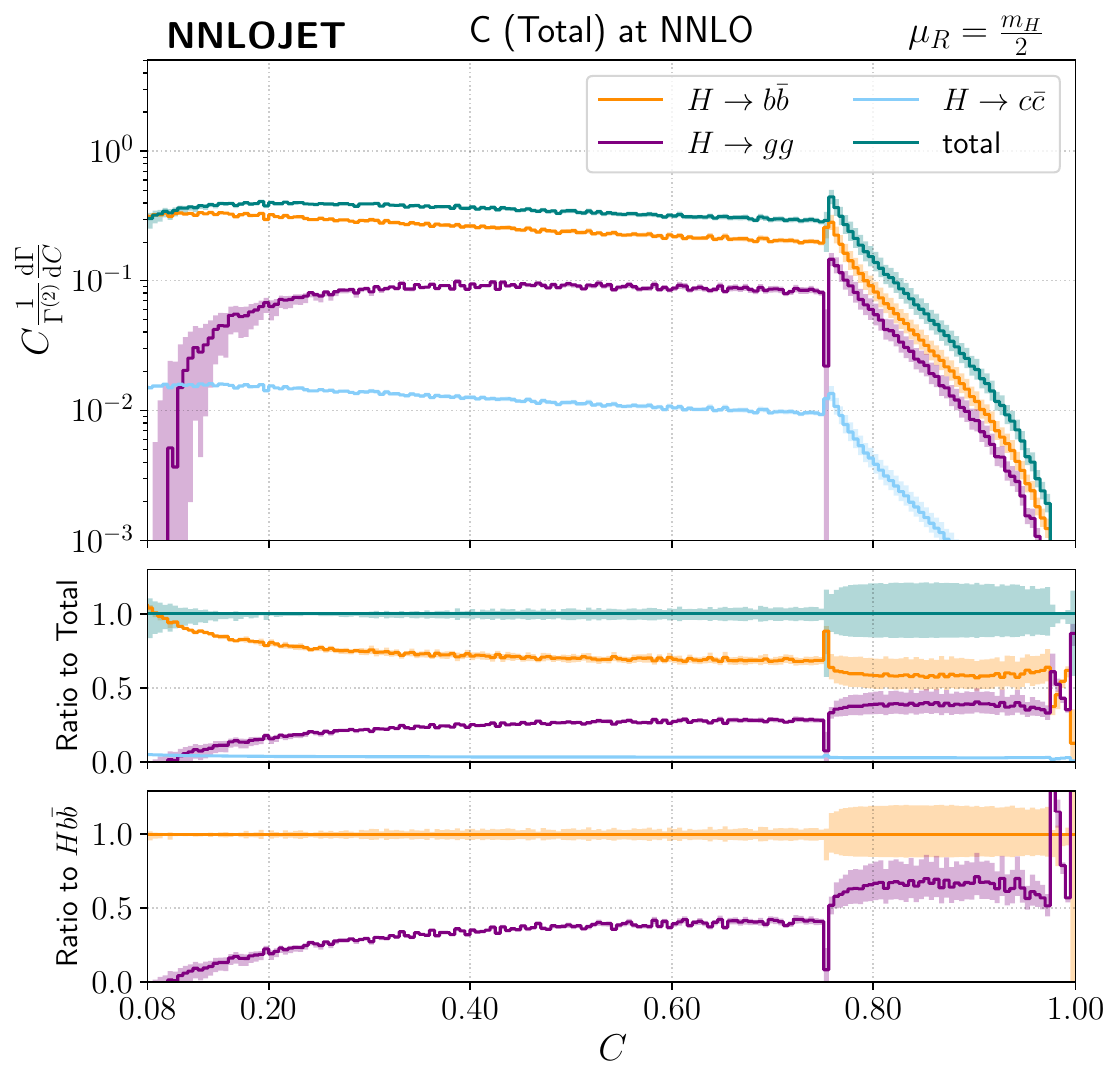}
  \includegraphics[width=0.45\textwidth]{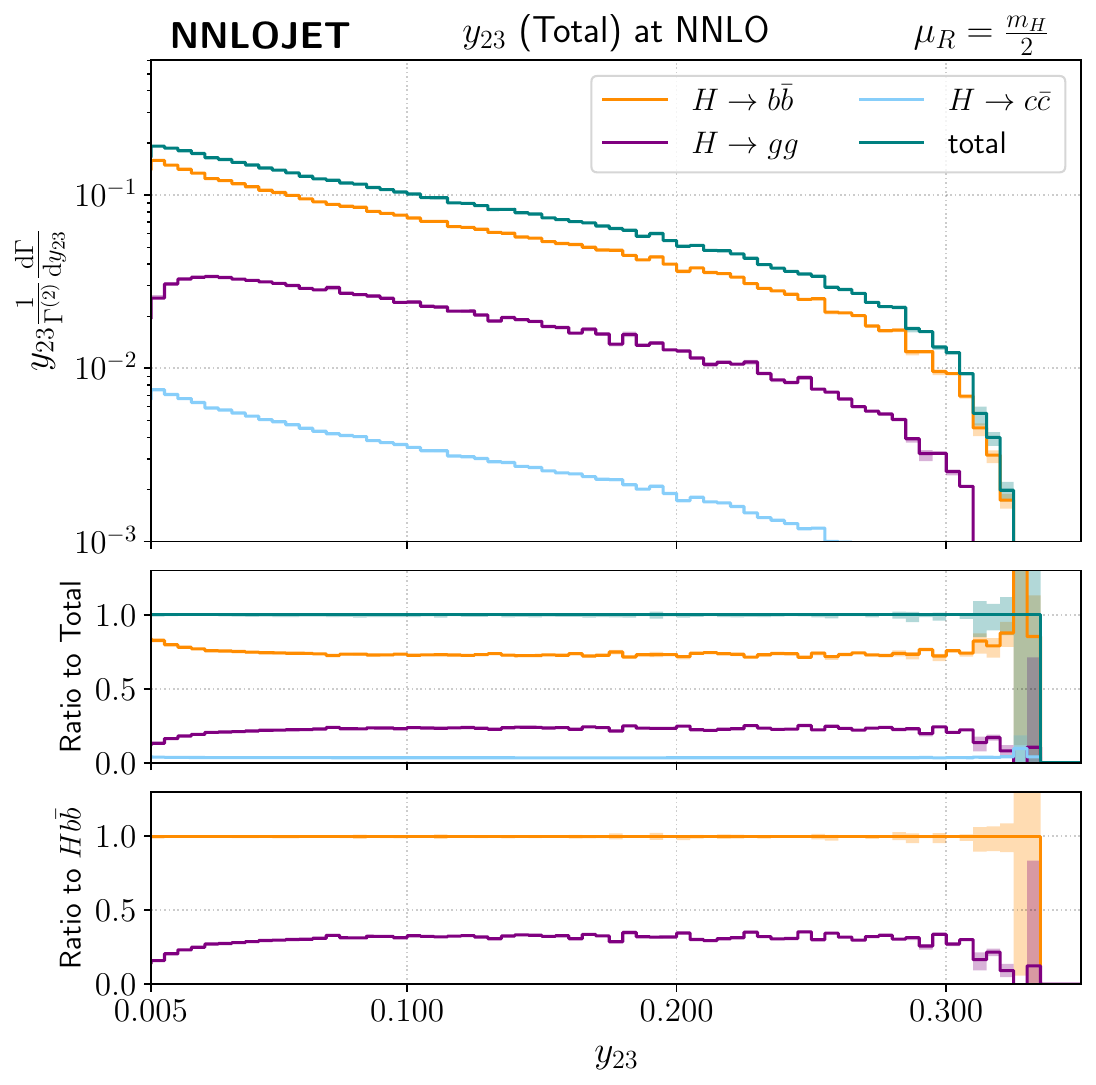}
  \caption{$C$-parameter (left column) and Durham $y_{23}$ (right column) distributions for the $H\to b\bar{b}$ channel (top row), the $H\to gg$ channel (middle row) and the sum of all decay modes at NNLO (bottom row). In the top and middle row, the LO (green), NLO (blue), and NNLO (red) are shown. In the bottom row, the individual contributions of the $H\to b\bar{b}$ (orange), $H\to gg$ (purple) and the $H\to c\bar{c}$ (light blue) channels are displayed alongside the sum (teal).}
  \label{fig:C_y23}
\end{figure}

In Figure~\ref{fig:C_y23}, we present the results for the $C$-parameter and the Durham $3$-jet resolution $y_{23}$. The distributions are truncated respectively at $C=0.08$ and $y_{23}=0.005$, below which the gluonic mode contributions turns negative.
The $C$-parameter has a Sudakov shoulder at $C=0.75$. It is actually this 
distribution that motivated the initial study of Sudakov shoulders in collider observables~\cite{Catani:1997xc},
when the kinematical range of a distribution changes with increasing multiplicity of the final-state radiation.  For the $C$-parameter distribution, the NNLO QCD corrections are positive and moderate in size (below
$20$\%) in 
the bulk region, which corresponds to $0.1<C<0.75$ in the Yukawa channel and to $0.3<C<0.75$ in the gluonic channel. 
In this region, perturbative convergence is decent, with NLO and NNLO uncertainty bands marginally 
overlapping. The residual uncertainty at NNLO is below $\pm5\%$ and $\pm7\%$ for the Yukawa and gluonic mode respectively. Below the 
bulk region, NNLO corrections turn negative and quickly increase in magnitude, thereby 
invalidating the fixed-order approach. Beyond the Sudakov shoulder at $C=0.75$, the NNLO corrections are large and
positive in both channels, almost doubling the NLO prediction, thus indicating a poor perturbative convergence. 
The relative contributions in the different channels resemble the behaviour in the  $\tau$-distribution: nearly constant 
fractions of about $75\%$ and $25\%$ in the Yukawa and gluonic channels in the bulk region, 
an increase of the relative importance of the gluonic channel above the Sudakov shoulder, and a decrease 
of the gluonic contribution towards the infrared region. 

Among all classical event-shape distributions, $y_{23}$ displays the smallest NNLO corrections 
in the bulk region, which is delimited by $0.005<y_{23}<0.25$ for both channels. In this region, the NNLO contributions are generally below $5\%$ of the NLO result in both channels, thereby showing excellent perturbative convergence and 
residual NNLO scale uncertainty at the level of $\pm2\%$ and $\pm3\%$ for the Yukawa and gluonic modes respectively.  
The NNLO corrections become larger 
and negative both below and above the bulk region, where perturbative convergence quickly deteriorates.
The relative contributions to the total distributions again amount to 
 about $75\%$ and $25\%$ from the Yukawa and gluonic channels.

\begin{figure}[htbp]
  \centering
  \includegraphics[width=0.32\textwidth]{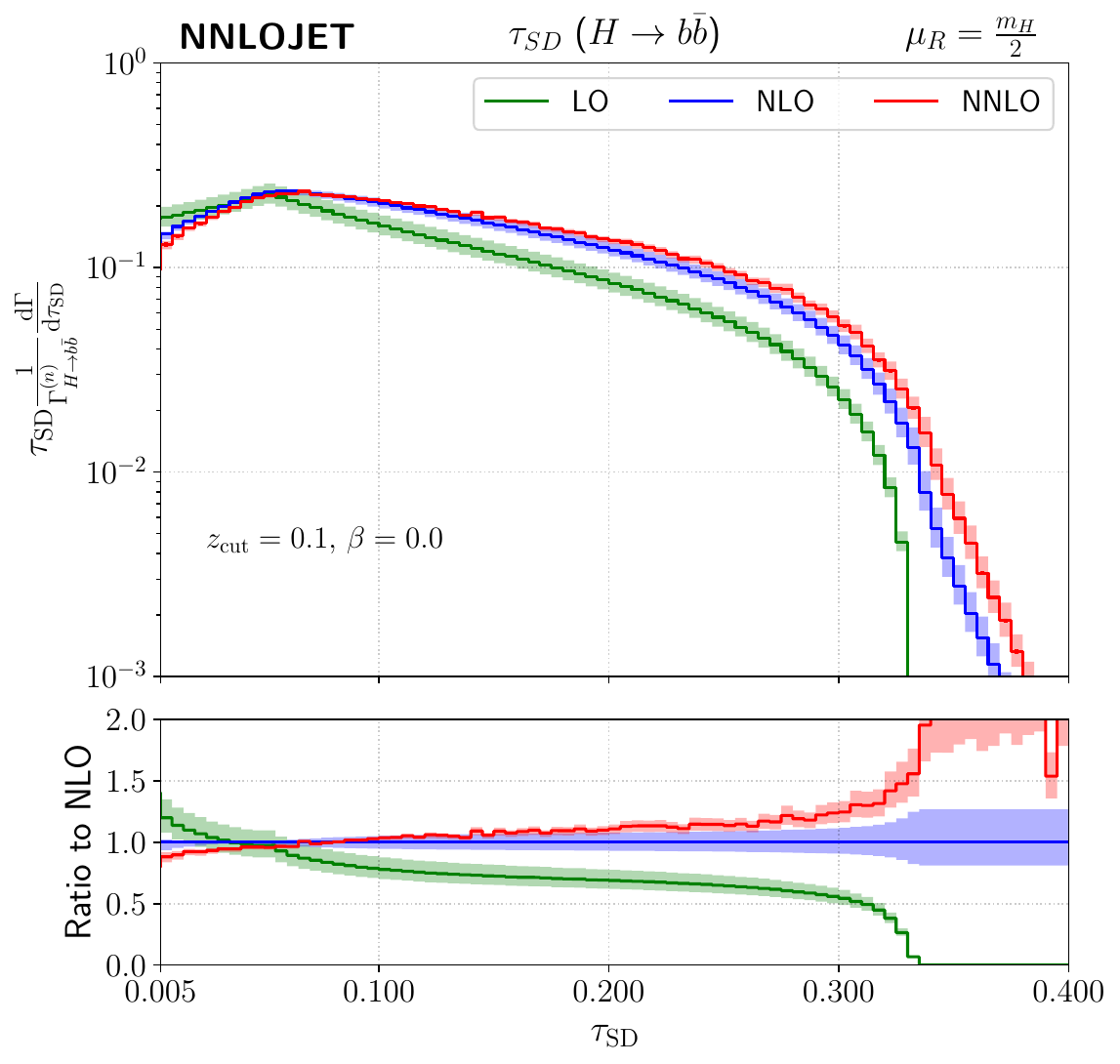}
  \includegraphics[width=0.32\textwidth]{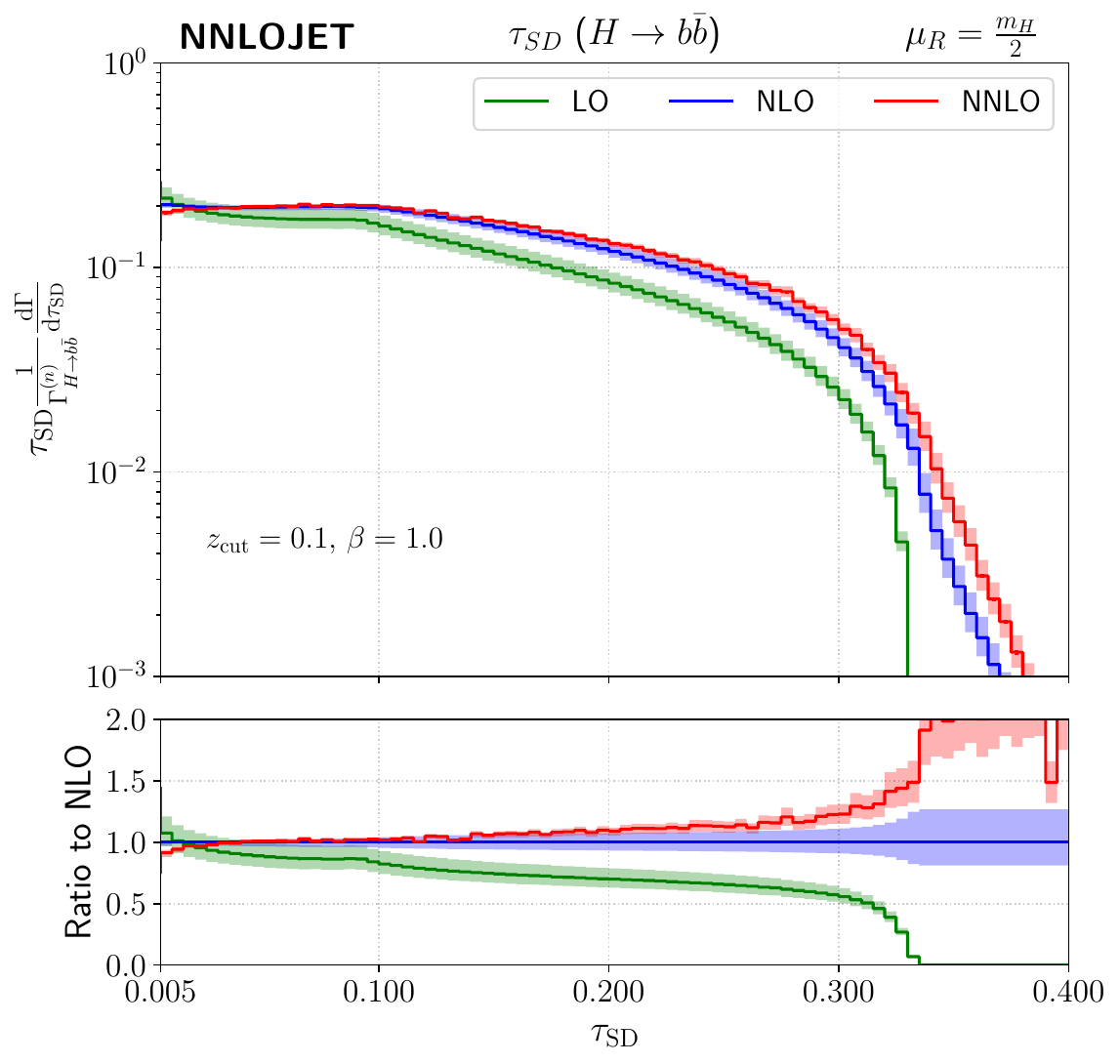}
  \includegraphics[width=0.32\textwidth]{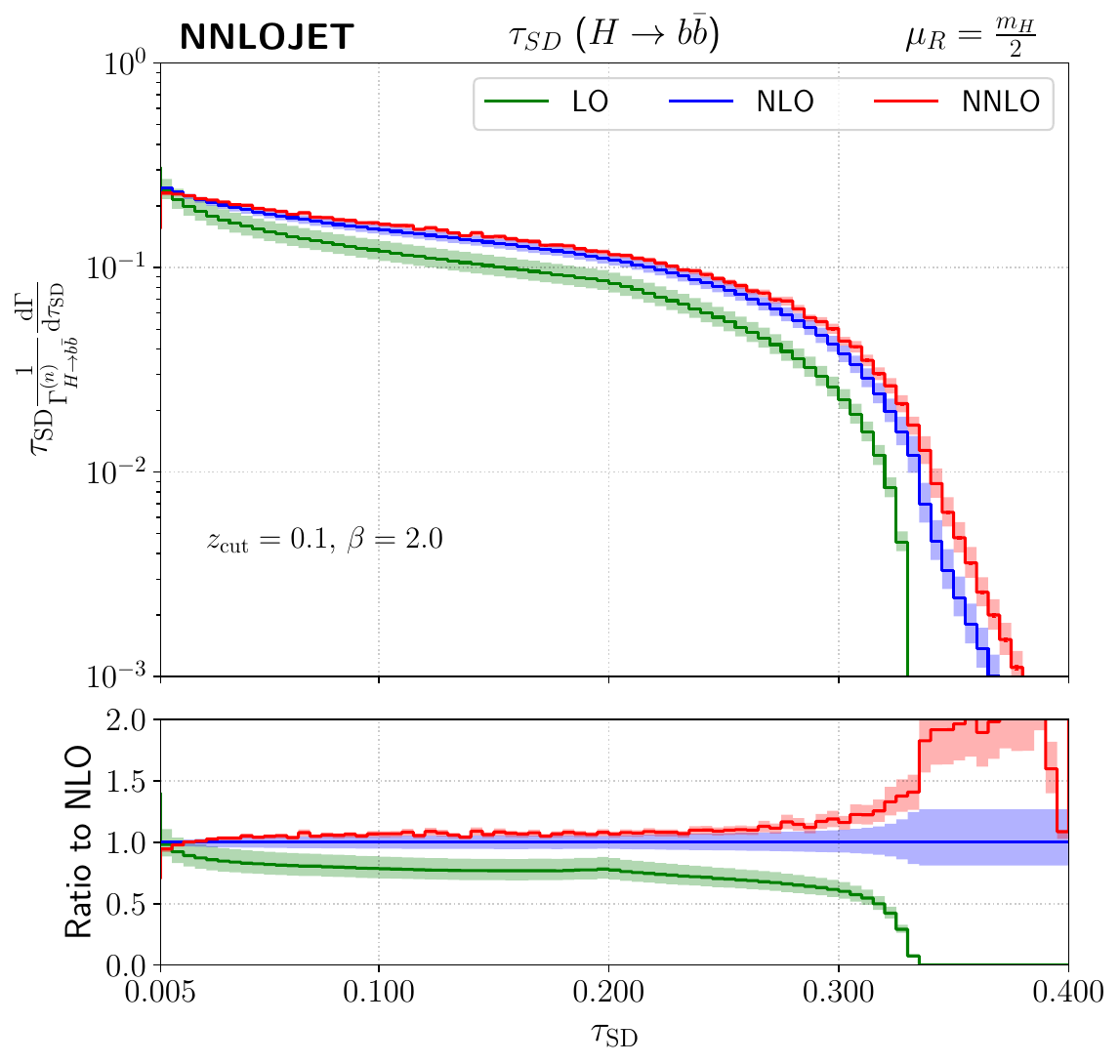}\\
  \includegraphics[width=0.32\textwidth]{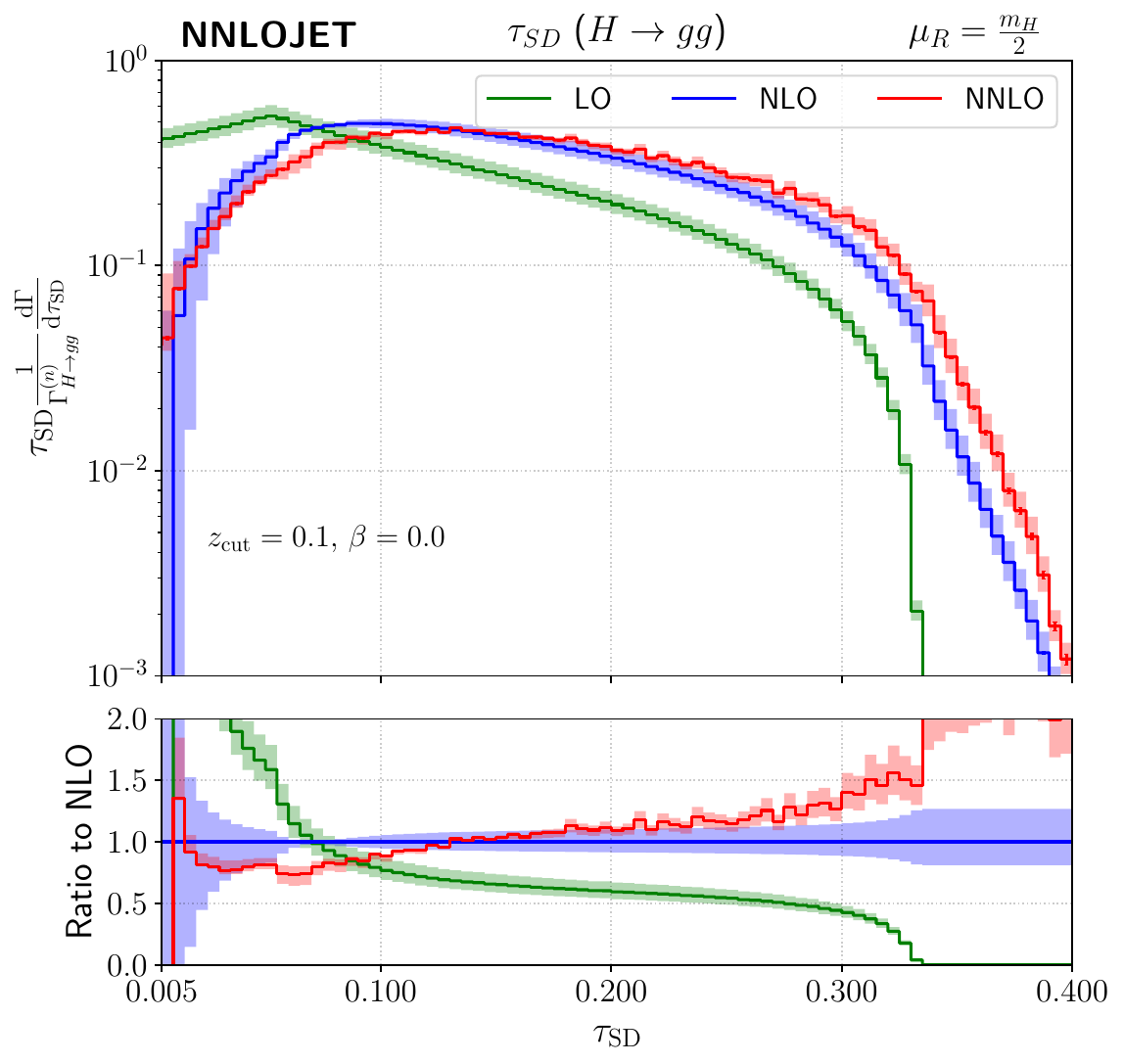}
  \includegraphics[width=0.32\textwidth]{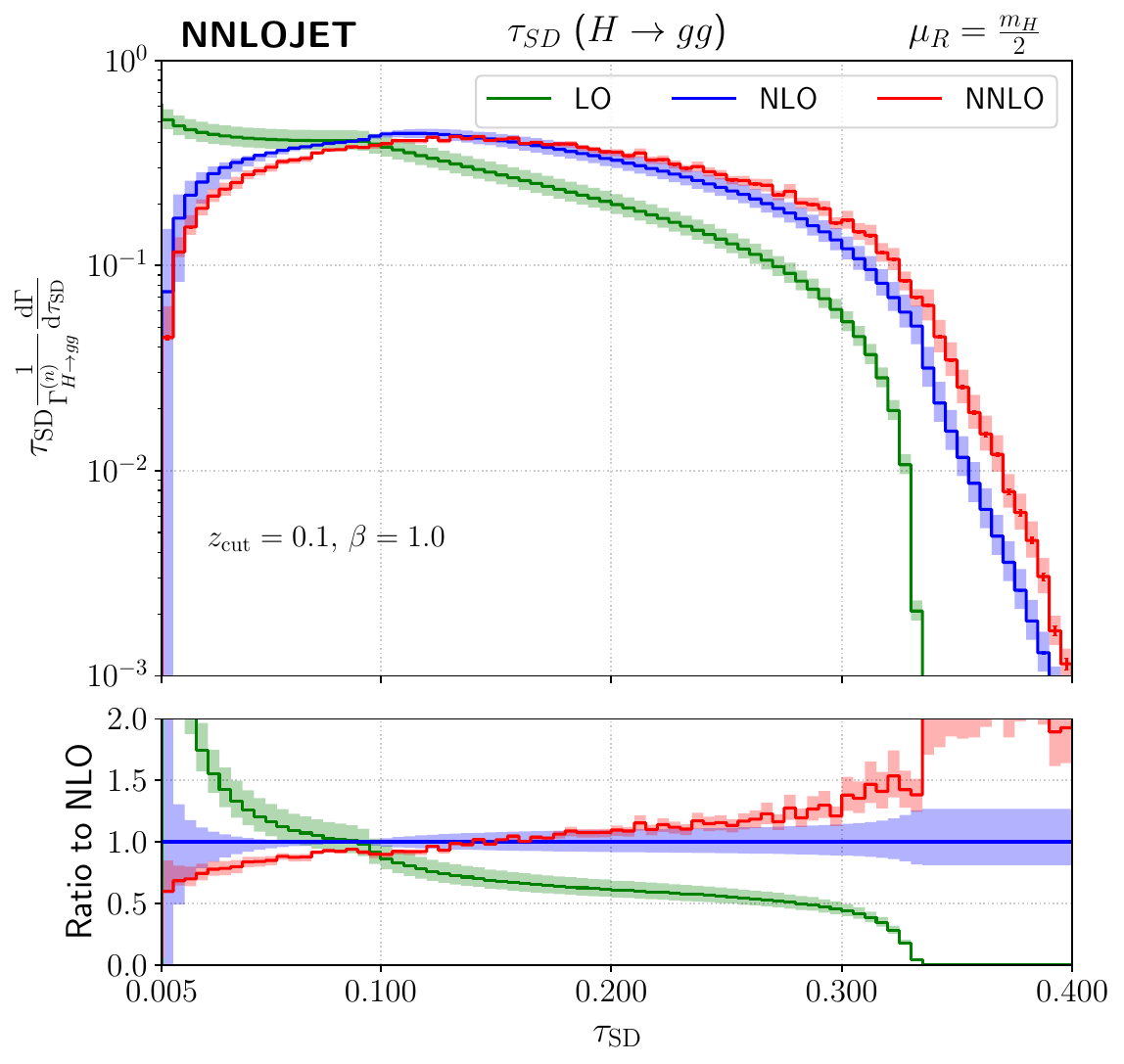}
  \includegraphics[width=0.32\textwidth]{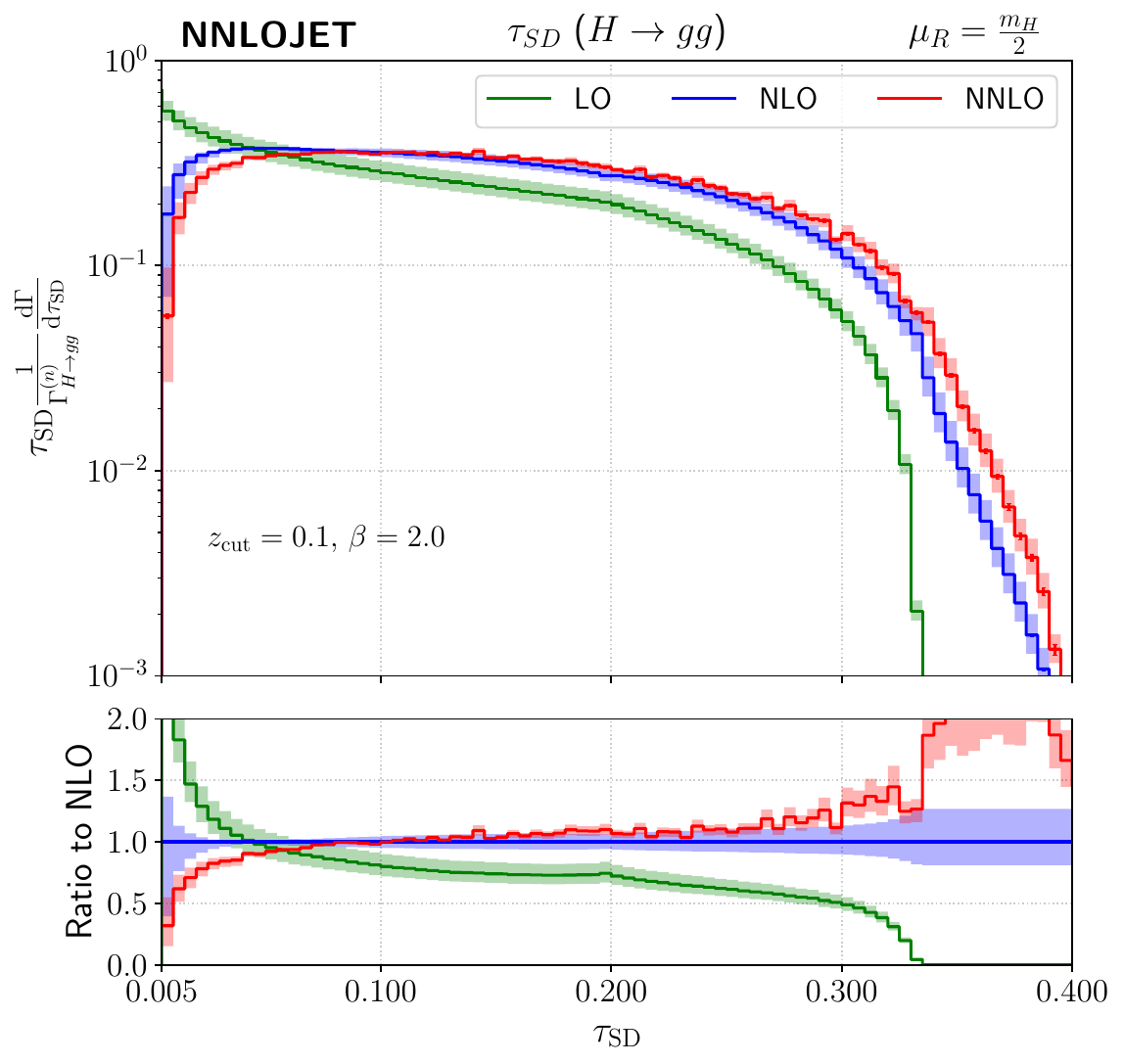}\\
  \includegraphics[width=0.32\textwidth]{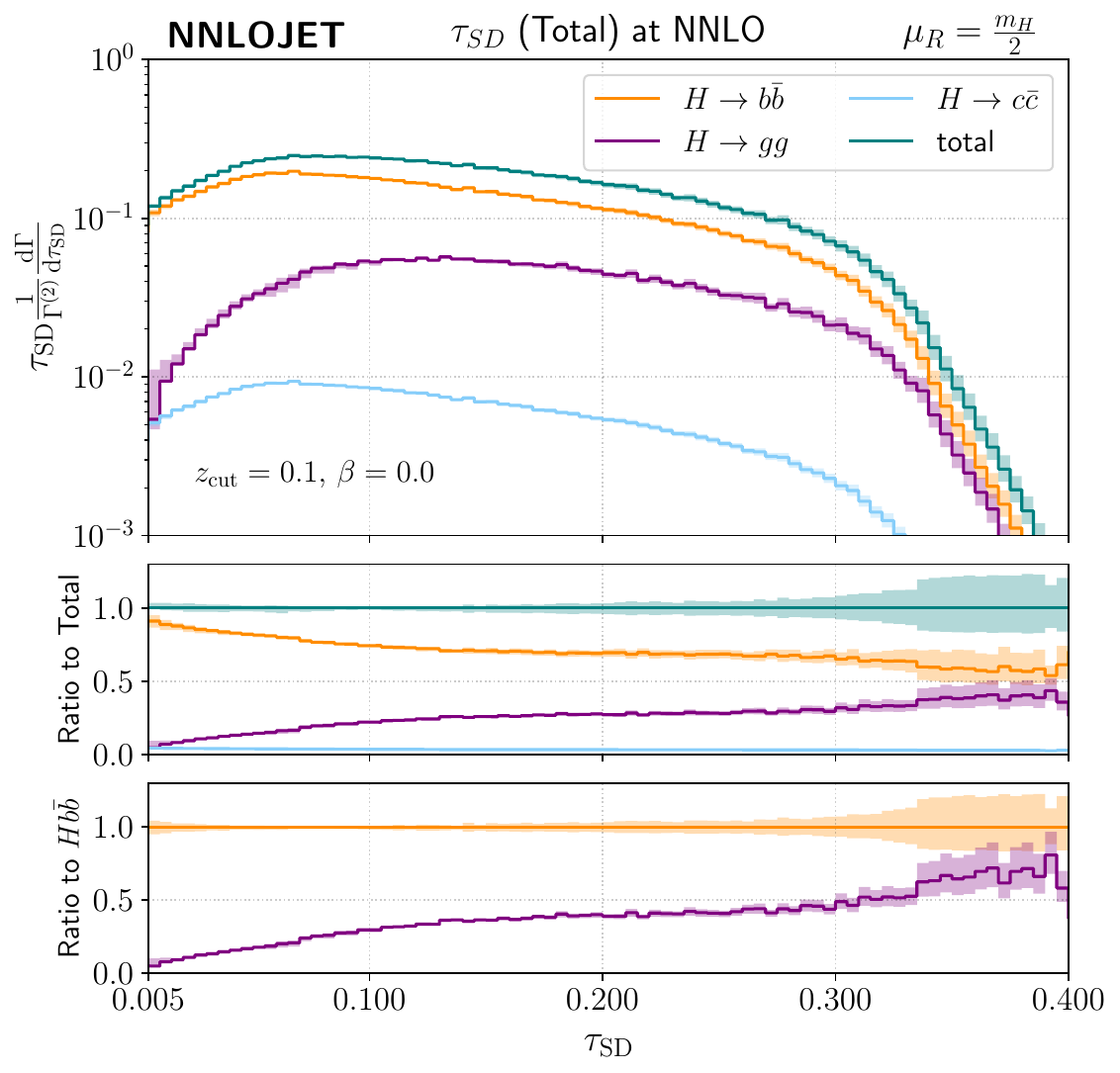}
  \includegraphics[width=0.32\textwidth]{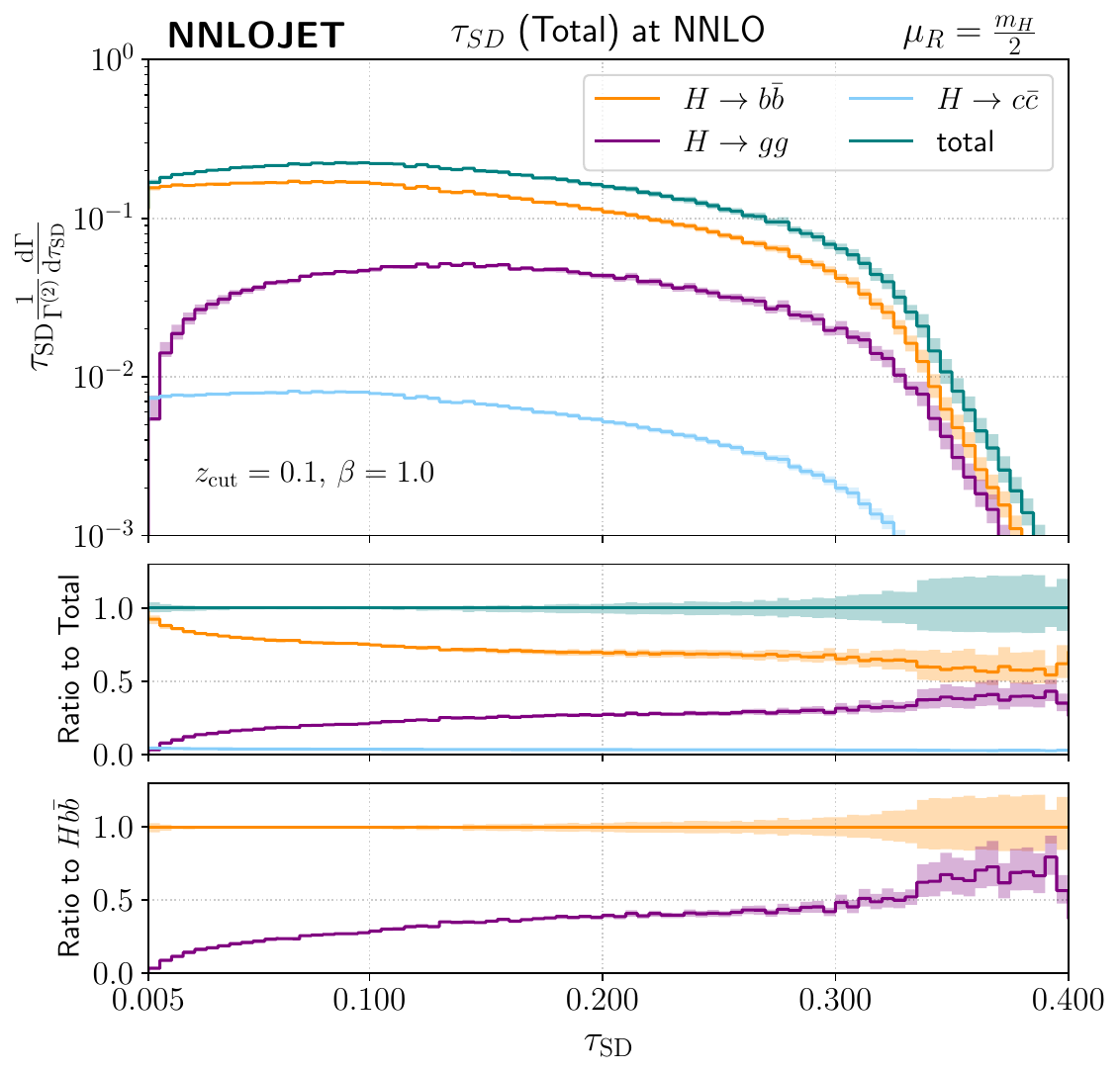}
  \includegraphics[width=0.32\textwidth]{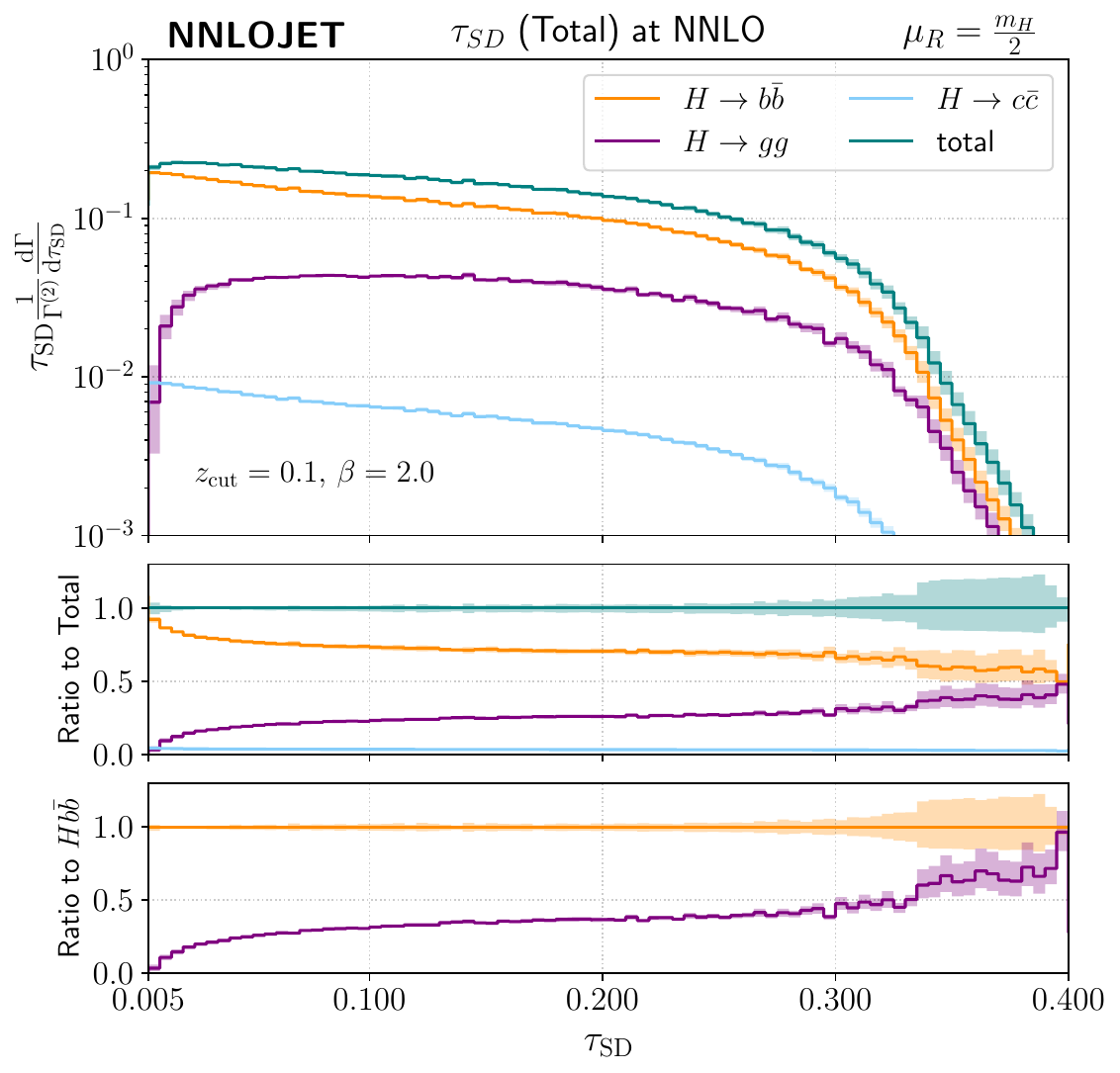}
  \caption{Soft-drop thrust distributions with $z_\mathrm{cut}=0.1$ and $\beta = 0$ (left column), $\beta = 1$ (middle columns), and $\beta = 2$ (right column) for the $H\to b\bar{b}$ channel (top row), the $H\to gg$ channel (middle row) and the sum of all decay modes at NNLO (bottom row). In the top and middle row, the LO (green), NLO (blue), and NNLO (red) are shown. In the bottom row, the individual contributions of the $H\to b\bar{b}$ (orange), $H\to gg$ (purple) and the $H\to c\bar{c}$ (light blue) channels are displayed alongside the sum (teal). }
  \label{fig:SDtau}
\end{figure}

In Figure~\ref{fig:SDtau}, we present the results for the soft-drop thrust $\tau_{\mathrm{SD}}$ for $z_\mathrm{cut}=0.1$ and three choices of the $\beta$ parameter: $\beta = 0,1,2$.
All distributions are truncated at $\tau_{\mathrm{SD}}=0.005$, below which the gluonic mode contributions turns negative. In both channels, 
the behaviour of the NNLO corrections in the bulk region and around the Sudakov shoulder closely resembles 
the $\tau$-distributions, Figure~\ref{fig:tau_HJM} (left). 
The introduction of the soft-drop criterion mainly affects the distributions in the approach to the infrared 
region, $\tau_{SD} \to 0$, where a substantial reduction of the NNLO corrections  
and consequently an improved perturbative convergence is observed as compared to 
the  $\tau$-distributions. The degree of improvement is strongest for $\beta=0$, it progressively decreases 
for $\beta=1$ and $\beta=2$, with the latter distributions approaching the $\tau$-distributions. The relative 
size of the Yukawa and gluonic modes to the total remain unchanged compared to the $\tau$-distributions for 
all values of $\beta$.

\newpage
\section{Conclusions}
\label{sec:conclusion}

We presented the first NNLO calculation of event-shape distributions in the hadronic decays of a Higgs boson. We consider the two dominant decay modes to bottom quarks via Yukawa interaction and to gluons via an effective vertex in the heavy top quark limit, and comment on the perturbative and kinematical features of the obtained distributions, as well as on the relevant differences between the decay modes. 

The size of the NNLO corrections varies among event shapes, but we observe in general mild NNLO corrections and  good perturbative convergence in the bulk of the distributions. Following the conventional choice in inclusive Higgs boson production, we consider $\mu_R=m_H/2$ as central renormalisation scale. As already observed in the calculation of jet rates~\cite{Fox:2025cuz}, the onset of large logarithmic corrections leading to a  perturbative breakdown in the limit of vanishing event-shape variables occurs earlier for the gluonic decay mode. We recover similar kinematical features to the ones present in event shapes in electron-positron annihilation. In particular, Sudakov shoulders arise where additional emissions populate previously forbidden phase space regions. 

Across all distributions, we observe that for intermediate and large event-shape values the relative size of the gluonic mode ($\approx25\%$) with respect to the sum of all decay channels is more than doubled with respect to its contribution in the inclusive decay of a Higgs boson ($\approx12\%$). The sensitivity to the gluonic decay mode is then enhanced in events with three well-resolved clusters, as already noticed in the study of the three-jet rate in~\cite{Fox:2025cuz}. Further enhancement is observed for large values of event-shape variables beyond their respective Sudakov shoulders, for example for thrust, total jet broadening and the $C$-parameter, where higher-multiplicity final states become relevant.

Our results pave the way for precision phenomenology in the hadronic decays of a Higgs boson at future electron-positron colliders. Moreover, they complement the rich literature on event shapes in electron-positron annihilation and offer important input for resummation and power correction studies for event shapes~\cite{Bhattacharya:2022dtm,Caola:2022vea,Bhattacharya:2023qet}, both in the case of fermionic and gluonic hard radiators. 

\acknowledgments
AG acknowledges the support of the Swiss National Science Foundation (SNF) under contract 200021-231259 and of the Swiss National Supercomputing Centre (CSCS) under project ID ETH5f. 
TG has received funding from the Swiss National Science Foundation (SNF)
under contract 200020-204200 and from the European Research Council (ERC) under
the European Union's Horizon 2020 research and innovation programme grant
agreement 101019620 (ERC Advanced Grant TOPUP).
NG gratefully acknowledges support from the UK Science and Technology Facilities Council (STFC) under contract ST/X000745/1 and hospitality from the Pauli Center for Theoretical Studies, Zurich. 
Part of the computations were carried out on the PLEIADES cluster at the University of Wuppertal, supported by the Deutsche Forschungsgemeinschaft (DFG, grant No. INST 218/78-1 FUGG) and the Bundesministerium f{\"u}r Bildung und Forschung (BMBF). MM is supported by a Royal Society Newton International Fellowship (NIF/R1/232539).

\appendix
\section{Inclusive Hadronic Decay Rates}\label{app:A}
The Higgs-boson inclusive  decay rates for the two channels we consider read
\begin{align}
  \Gamma^{(k)}_{H\to b\bar{b}} &= \Gamma^{(0)}_{H\to b\bar{b}}\, \left(1+\sum\limits_{n=1}^k\left(\dfrac{\alphas}{2\pi}\right)^n C_{b\bar{b}}^{(n)}\right) \,,\\
  \Gamma^{(k)}_{H\to gg} &= \Gamma^{(0)}_{H\to gg}\, \left(1+\sum\limits_{n=1}^k\left(\dfrac{\alphas}{2\pi}\right)^n C_{gg}^{(n)}\right) \,.
  \label{eq:ratesNkLO1}
\end{align}
The perturbative corrections~\cite{Herzog:2017dtz}  are given at \NLO by
\begin{align}
  C_{b\bar{b}}^{(1)} &=  \dfrac{r_1}{2}+2\gamma_0 L_R\,,\\
  C_{gg}^{(1)} &= c_1+\dfrac{1}{2}g_1+2\beta_0 L_R\,,
\end{align}
and at NNLO by
\begin{align}
  C_{b\bar{b}}^{(2)} &=\dfrac{1}{4}\left(r_2+(8\gamma_1+4r_1\gamma_0+2r_1\beta_0)L_R+(8\gamma_0^2+4\beta_0\gamma_0)L_R^2\right)\,,\\
  C_{gg}^{(2)} &=  \dfrac{1}{4}\left(2c_2+c_1^2+g_2+(16\beta_1+6\beta_0g_1)L_R+12\beta_0^2L_R^2+2c_1(g_1+4\beta_0L_R)\right)\,,
\end{align}
where
\begin{equation}
  L_R = \log\frac{\mu_R^2}{m_H^2}.
\end{equation}
The running of the strong and Yukawa coupling is defined by
\begin{equation}
  \begin{split}
    \dfrac{d\alpha_s}{d\log(\mu_R^2)} &=\mu_R^2\dfrac{d\alpha_s}{d\mu_R^2}=-\alpha_s\left(\beta_0\left(\dfrac{\alpha_s}{2\pi}\right)+\beta_1\left(\dfrac{\alpha_s}{2\pi}\right)^2+\mathcal{O}(\alpha_s^3)\right)\,,\\
    \dfrac{dy}{d\log(\mu_R^2)} &=\mu_R^2\dfrac{dy}{d\mu_R^2}=-y\left(\gamma_0\left(\dfrac{\alpha_s}{2\pi}\right)+\gamma_1\left(\dfrac{\alpha_s}{2\pi}\right)^2+\mathcal{O}(\alpha_s^3)\right)\,,
  \end{split}
\end{equation}
with
\begin{equation}
  \begin{split}
    \beta_0 &= \dfrac{11}{6}C_A-\dfrac{2}{3}T_FN_F\,,\\
    \beta_1 &= \dfrac{17}{6}C_A^2-\dfrac{5}{3}C_AT_FN_F-C_FT_FN_F\,,\\
    \gamma_0 &= \dfrac{3}{2}C_F\,,\\
    \gamma_1 &= \dfrac{1}{4}\left(\dfrac{3}{2}C_F^2+\dfrac{97}{6}C_FC_A-\dfrac{10}{3}C_FT_FN_F\right)\,.
  \end{split}
\end{equation}
The numerical constants above are
\begin{equation}
  \begin{split}
      r_1 &= 17C_F \,,\\
      g_1 &= \dfrac{73}{3}C_A-\dfrac{14}{3}N_F \,,\\
      c_1 &= 11\,,\\
      r_2 &= C_F^2\left(\dfrac{691}{4}-36\zeta_2-36\zeta_3\right)+C_AC_F\left(\dfrac{893}{4}-22\zeta_2-62\zeta_3\right)\,,\\
      & -C_FN_F\left(\dfrac{65}{2}-4\zeta_2-8\zeta_3\right)\,,\\
      g_2 &=  C_A^2\left(\dfrac{37631}{54}-\dfrac{242}{3}\zeta_2-110\zeta_3\right)-C_AN_F\left(\dfrac{6665}{27}-\dfrac{88}{3}\zeta_2+4\zeta_3\right)\,,\\
      & -C_FN_F\left(\dfrac{131}{3}-24\zeta_3\right)+N_F^2\left(\dfrac{508}{27}-\dfrac{8}{3}\zeta_2\right)\,,\\
      c_2 &= \dfrac{2777}{18}+19\log\left(\dfrac{\mu_R^2}{m_t^2}\right)-N_F\left(\dfrac{67}{6}-\dfrac{16}{3}\log\left(\dfrac{\mu_R^2}{m_t^2}\right)\right)\,.
  \end{split}
\end{equation}

\section{Renormalisation Scale Dependence}\label{app:B}
For the Yukawa  channel, the renormalisation scale dependence of the expansion coefficients defined in (\ref{eq:exp1}) is given by
\begin{equation}
  \begin{split}
    A_{b\bar{b}}(\mu_R) &= A_{b\bar{b}}(\mu_0) \,, \\
    B_{b\bar{b}}(\mu_R) &= B_{b\bar{b}}(\mu_0)+(\beta_0+2\gamma_0)L_RA_{b\bar{b}}(\mu_0) \,, \\
    C_{b\bar{b}}(\mu_R) &= C_{b\bar{b}}(\mu_0)+(2\beta_0+2\gamma_0)L_RB_{b\bar{b}}(\mu_0) \\
    & +\left((\beta_1+2\gamma_1)L_R+(\beta_0+\gamma_0)(\beta_0+2\gamma_0)L_R^2)\right)B_{b\bar{b}}(\mu_0) \,.
  \end{split}
\end{equation}
where
\begin{equation}
L _R= \log\frac{\mu_R^2}{\mu_0^2}.
\end{equation}
For the gluonic channel:
\begin{equation}
  \begin{split}
  A_{gg}(\mu_R) &= A_{gg}(\mu_0)\,,\\
  B_{gg}(\mu_R) &= B_{gg}(\mu_0)+3\beta_0L_RA_{gg}(\mu_0)\,,\\
  C_{gg}(\mu_R) &= C_{gg}(\mu_0)+4\beta_0L_RB_{gg}(\mu_0)\,,\\
  &+\left(\beta_1L_R+6\beta_0^2L_R^2-(C_1\beta_0-2\beta_1)L_t\right)A_{gg}(\mu_0)\,,
  \end{split}
\end{equation}
where
\begin{equation}
  L_t = \log\left(\dfrac{m_t(\mu_R)^2}{m_t(\mu_0)^2}\right).
\end{equation}
The analogous formulae for the expansion coefficients in~\eqref{eq:exp2} can be obtained by combining the above with the results in Appendix~\ref{app:A}.

\bibliography{bibliography.bib}

\end{document}